\documentclass[a4paper,11pt]{article}
\pdfoutput=1
\usepackage{jheppub}
\usepackage[T1]{fontenc}
\usepackage{subfigure}
\usepackage{overpic}
\usepackage{multirow}

\newcommand{\Kbar}{\bar{K}^{0}}
\newcommand{\Kstar}{K^{*}(892)^{0}}
\newcommand{\Kstarc}{K^{*}(892)^{+}}
\newcommand{\Kstarcm}{K^{*}(892)^{-}}

\newcommand{\KstarT}{K^{*}_{2}(1430)^{0}}

\newcommand{\KstarcT}{K^{*}_{2}(1430)^{+}}

\newcommand{\Ks}{K_{S}^{0}}
\newcommand{\Kl}{K_{L}^{0}}

\title{\boldmath Measurement of the $e^{+}e^{-} \to \Ks \Kl \pi^{0}$ cross sections from $\sqrt{s}=$ 2.000 to 3.080 GeV}

\collaborationImg{\includegraphics[height=3cm,angle=90]{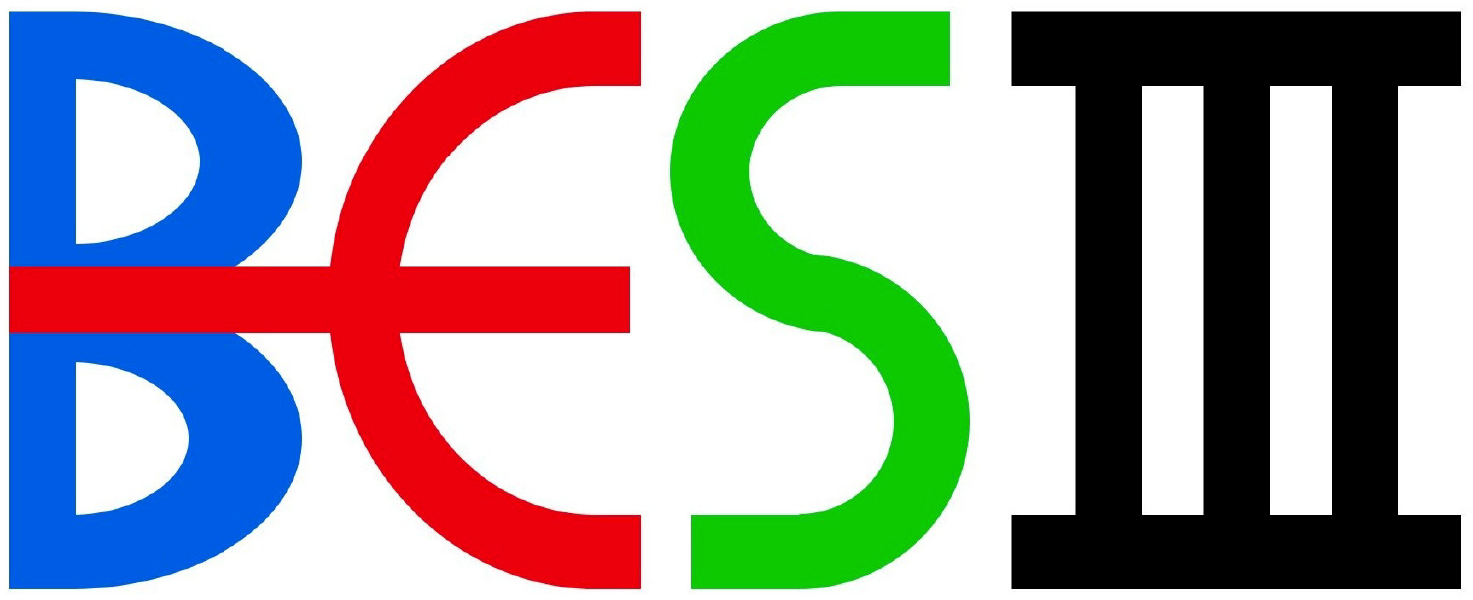}}
\collaboration{The BESIII Collaboration}

\emailAdd{besiii-publications@ihep.ac.cn}

\abstract{Based on $e^{+}e^{-}$ collision data collected at
  center-of-mass energies from 2.000 to 3.080 GeV by the BESIII
  detector at the BEPCII collider, a partial wave analysis is
  performed for the process $e^{+}e^{-}\to \Ks \Kl \pi^{0}$.  The
  results allow the Born cross sections of the process $e^{+}e^{-}\to
  \Ks \Kl \pi^{0}$, as well as its subprocesses $e^{+}e^{-}\to
  \Kstar\Kbar$ and $\KstarT\Kbar$ to be measured. The Born cross
  sections for $e^{+}e^{-}\to\Ks\Kl\pi^{0}$ are consistent with
  previous measurements by BaBar, but with substantially improved
  precision. The Born cross section lineshape of the process
  $e^{+}e^{-}\to\Kstar\Kbar$ is consistent with a vector meson state
  around 2.2 GeV with a significance of 3.2$\sigma$. A
  Breit-Wigner fit determines its mass as
  $M_Y=(2164.7\pm9.1\pm3.1)~{\rm{MeV}}/c^{2}$ and its width as
  $\Gamma_{Y}=(32.4\pm21.0\pm1.8)~\rm{MeV}$.}

  
\arxivnumber{....}

\begin{document} \maketitle \flushbottom
\section{Introduction} 

The vector meson state $Y(2175)$, denoted as $\phi(2170)$ by the
Particle Data Group~\cite{PDG}, is one of the more interesting
particles in the field of light-flavored hadron spectroscopy. It was
first observed by the BaBar collaboration~\cite{babar_phif0} and
subsequently investigated by the Belle, BES and BESIII
collaborations~\cite{belle_isr_phi_pipi,bes_etaphif0,bes_etaphipipi,bes_etaY2175,bes_phieta,bes_phietap,bes_phipipi,besiii_kk,bes_kskl,besiii_kkpi,besiii_kkpipi}. Several
interpretations have been proposed for the $\phi(2170)$ state, such as
a conventional $3^3S_1$ or $2^3D_1$~$s\bar{s}$
state~\cite{ssbar_1,ssbar_2,ssbar_3,ssbar_4}, a $s\bar{s}g$
hybrid~\cite{hybrid_1,hybrid_2}, a tetraquark
state~\cite{trq_1,trq_2,trq_3,trq_4}, a $\Lambda\bar{\Lambda}(^3S_1)$
bound state~\cite{llb_1,llb_2,llb_3}, and a $\phi K\bar{K}$ resonance
state~\cite{phikk_1}.

Studies of the $\phi(2170)$ have been carried out using various final
states such as $\phi\eta$~\cite{bes_phieta,babar_kskpi},
$\phi\eta^{\prime}$~\cite{bes_phietap}, $\phi
f_{0}(980)$~\cite{babar_phif0,babar_isr_kkpipi,belle_isr_phi_pipi,bes_etaphif0,bes_etaphipipi,bes_etaY2175,bes_phipipi},
$K^{+}K^{-}$~\cite{besiii_kk,babar_kk,babar_kk_sim},
$\Ks\Kl$~\cite{bes_kskl,babar_kskl}, $\Kstarc
K^{-}$~\cite{besiii_kkpi}, $\KstarcT K^{-}$~\cite{besiii_kkpi},
$\Kstarc\Kstarcm$~\cite{besiii_kkpipi} and other charged excited $K\bar{K}$
states~\cite{besiii_kkpipi}. None of these final states is dominant,
and the product of the $e^{+}e^{-}$ partial width and the branching
fraction~(BF) of each final state is consistently below 100 eV. The
partial decay width 
$\Gamma(\phi\eta)$~\cite{bes_phieta,babar_kskpi} is less than 
$\Gamma(\phi\eta^{\prime})$~\cite{bes_phietap}, which disfavors the
hybrid interpretation~\cite{hybrid_1,hybrid_2}. This result can
be explained by the hadronic transition of a strangeonium-like meson
along with $\eta-\eta^{\prime}$ mixing~\cite{Mayunheng}. The partial
width of $\Kstarc\Kstarcm$ is significantly greater than that of
$K_{1}(1400)^{+}K^{-}$, as predicted for the $2^3D_1$ or $3^3S_1$
state~\cite{ssbar_1,ssbar_2,ssbar_3,ssbar_4}. However, the BESIII
collaboration has observed a clear structure in the cross section line
shape of $K_{1}(1400)^{+}K^{-}$ around 2.2 GeV~\cite{besiii_kkpipi},
but no enhancement in the cross section line shape for
$\Kstarc\Kstarcm$~\cite{besiii_kkpipi}, which disfavors the $2^3D_1$
or $3^3S_1$ prediction. The BESIII collaboration has also measured a
larger partial width of $\KstarcT K^{-}$ compared to $\Kstarc K^{-}$
at center-of-mass (c.m.) energies~($\sqrt{s}$) from 2.000 to 3.080
GeV~\cite{besiii_kkpi}, which contradicts the prediction that the
$\phi(2170)$ is the $2^3D_1$ strangeonium state~\cite{ssbar_3}.  More
precise measurements of the decay properties of the $\phi(2170)$ are
desirable to better understand the nature of the $\phi(2170)$.
    
The $e^{+}e^{-}\to\Ks\Kl\pi^{0}$~\cite{babar_ksklpi} and
$e^{+}e^{-}\to \Ks K^{\pm}\pi^{\mp}$~\cite{babar_kskpi} processes have
been investigated by the BaBar collaboration using the initial state
radiation~(ISR) technique. A Dalitz amplitude analysis was performed
for $e^{+}e^{-}\to \Ks K^{\pm}\pi^{\mp}$, leading to the determination
of the isoscalar and isovector cross sections for $K^{*}(892)\bar{K}$. A
distinct asymmetry between neutral and charged channels is observed in
the Dalitz plot for $\Ks\pi^{\mp}$ and $K^{\pm}\pi^{\mp}$ within
$\sqrt{s^{\prime}}=2-3$ GeV. It may be related to a similar effect observed in
the radiative decay rates of the neutral and charged
$K^{*}_{2}(1430)$~\cite{babar_kskpi}. The SND collaboration has
studied $e^{+}e^{-}\to\Ks\Kl\pi^{0}$ at $\sqrt{s}=1.3-2.0$ GeV, and
the cross sections have been measured at a statistical uncertainty
level of 10\%-30\%~\cite{snd_ksklpi}.

In this paper, we present a partial wave analysis~(PWA) of the process
$e^{+}e^{-}\to\Ks\Kl\pi^{0}$ based on 19 data samples collected by the
BESIII experiment, ranging from $\sqrt{s} = 2.000$ to 3.080 GeV and
corresponding to an integrated luminosity of 647
$\rm{pb}^{-1}$~\cite{besiii_lumin,bes_omegapippim}. The Born cross
section of the process $e^{+}e^{-}\to \Ks \Kl \pi^{0}$ and its
sub-processes $e^{+}e^{-}\to \Kstar\Kbar$ and $\KstarT\Kbar$ are
measured. Throughout the paper charge conjugated processes are also
included by default.

\section{BESIII detector and Monte Carlo simulation}

The BESIII detector~\cite{besiii} records symmetric $e^{+}e^{-}$
collisions provided by the BEPCII storage ring~\cite{bepcii}, which
operates with a peak luminosity of
$1\times10^{33}~\rm{cm}^{-2}\rm{s}^{-1}$ in the range of $\sqrt{s}$
from 2.0 to 4.95 GeV. BESIII has collected large data samples in this
energy region~\cite{BESIIIDATA}. The cylindrical core of the BESIII
detector covers 93\% of the full solid angle and consists of a
helium-based multilayer drift chamber~(MDC), a plastic scintillator
time-of-flight system~(TOF), and a CsI(Tl) electromagnetic
calorimeter~(EMC), which are all enclosed in a superconducting
solenoidal magnet providing a 1.0 T~(0.9 T in 2012) magnetic
field. The solenoid is supported by an octagonal flux-return yoke with
resistive plate counter muon identification modules interleaved with
steel. The charged-particle momentum resolution at 1 GeV$/c$ is 0.5\%,
and the d$E$/d$x$ resolution is 6\% for electrons from Bhabha
scattering. The EMC measures photon energies with a resolution of
2.5\%~(5\%) at 1 GeV in the barrel~(end cap) region. The time
resolution in the TOF barrel region is 68 ps, while that in the end
cap region is 110 ps.

Simulated samples produced with {\sc{GEANT4}} based~\cite{bes_g4}
Monte Carlo~(MC) software, which includes the geometric
description~\cite{detvis} of the BESIII detector and the detector
response, are used to optimize the event selection criteria, estimate
backgrounds, and determine the detection efficiency. The signal MC
samples for the processes $e^{+}e^{-}\to \Ks\Kl\pi^{0}$, $\Kstar\Kbar$
and $\KstarT\Kbar$ are generated by
{\sc{ConExc}}~\cite{bes_conexc} using
an amplitude model with parameters fixed to the PWA results. For
background studies, inclusive hadronic events are generated with a
hybrid generator that includes {\sc{ConExc}},
{\sc{LUARLW}}~\cite{bes_luarlw} and
{\sc{PHOKHARA}}~\cite{bes_phokhara}.

\section{Event selection and background analysis}

The signal process $e^{+}e^{-} \to \Ks \Kl \pi^{0}$ is reconstructed
with $\Ks\to\pi^{+}\pi^{-}$, $\pi^{0}\to\gamma\gamma$, and $\Kl$
treated as a missing particle. Signal candidates are required to
have two charged pions with zero net charge and at least two photons.

Charged tracks detected in the MDC are required to be within a polar
angle ($\theta$) range of $|\rm{\cos\theta}|<0.93$. Here, $\theta$ is
defined with respect to the $z$-axis, which is the symmetry axis of
the MDC. Each $\Ks$ candidate is reconstructed from two oppositely
charged tracks satisfying that the distance of closest approach to the
interaction point~(IP) must be less than 20~cm along the $z$-axis. The
two charged tracks are assigned as $\pi^+\pi^-$ without imposing
further particle identification criteria. They are constrained to
originate from a common vertex and are required to have an invariant
mass within $|M(\pi^{+}\pi^{-}) - m_{\Ks}|<$ 12~MeV$/c^{2}$, where $M(\pi^{+}\pi^{-})$ is the invariant mass of $\pi^{+}\pi^{-}$ pair with kinematics updated by the vertex fit and 
$m_{\Ks}$ is the $\Ks$ nominal mass~\cite{PDG}. The decay length of
the $\Ks$ candidate is required to be greater than twice the vertex
resolution away from the IP.

Photon candidates are identified using showers in the EMC. The
deposited energy of each shower must be more than 25 MeV in the barrel
region~($|\!\cos\theta|<0.80$) and more than 50 MeV in the end cap
region~($0.86<|\!\cos\theta|<0.92$). To exclude showers that originate
from charged tracks, the angle subtended by the EMC shower and the
position of the closest charged track at the EMC must be greater than
$10^\circ$ as measured from the IP. To suppress electronic noise and
showers unrelated to the event, the difference between the event start
time and the EMC time of the photon candidate is required to be within
\mbox{[0,~700] ns}.

To suppress background and improve the kinematic resolution, a
one-constraint~(1C) kinematic fit imposing energy-momentum
conservation is carried out under the $\Ks\Kl\gamma \gamma$ hypothesis
with $\Kl$ treated as a missing particle. If there are more than two
photons in an event, the combination with the minimum
$\chi_{\rm{1C}}^{2}$ is retained for further analysis, and candidate
events are required to satisfy $\chi_{\rm{1C}}^{2}<30$. To suppress
the contamination from the process $e^{+}e^{-}\to
\gamma_{\rm{ISR}}\Ks\Kl$, an additional 1C kinematic fit is performed
under the hypothesis of $\gamma\Ks\Kl$, and only events which satisfy
$\chi_{\rm{1C}}^{2}<\chi_{\rm{1C}}^{2}(\gamma\Ks\Kl)$ are retained. To
remove $\Kl$ showers in the EMC that could be mistaken as photons, the
angles between the candidate EMC shower and the $\Kl$ momentum after
the kinematic fit are required to be greater than 20$^{\circ}$. Each
signal candidate is required to have the invariant mass of the two
photons within the $\pi^{0}$ mass
region~(\mbox{$|M(\gamma\gamma)-m_{\pi^{0}}|<0.015$ GeV/$c^{2}$}).

\begin{figure}[t]
        \centering
        \begin{overpic}[width=0.48\textwidth]{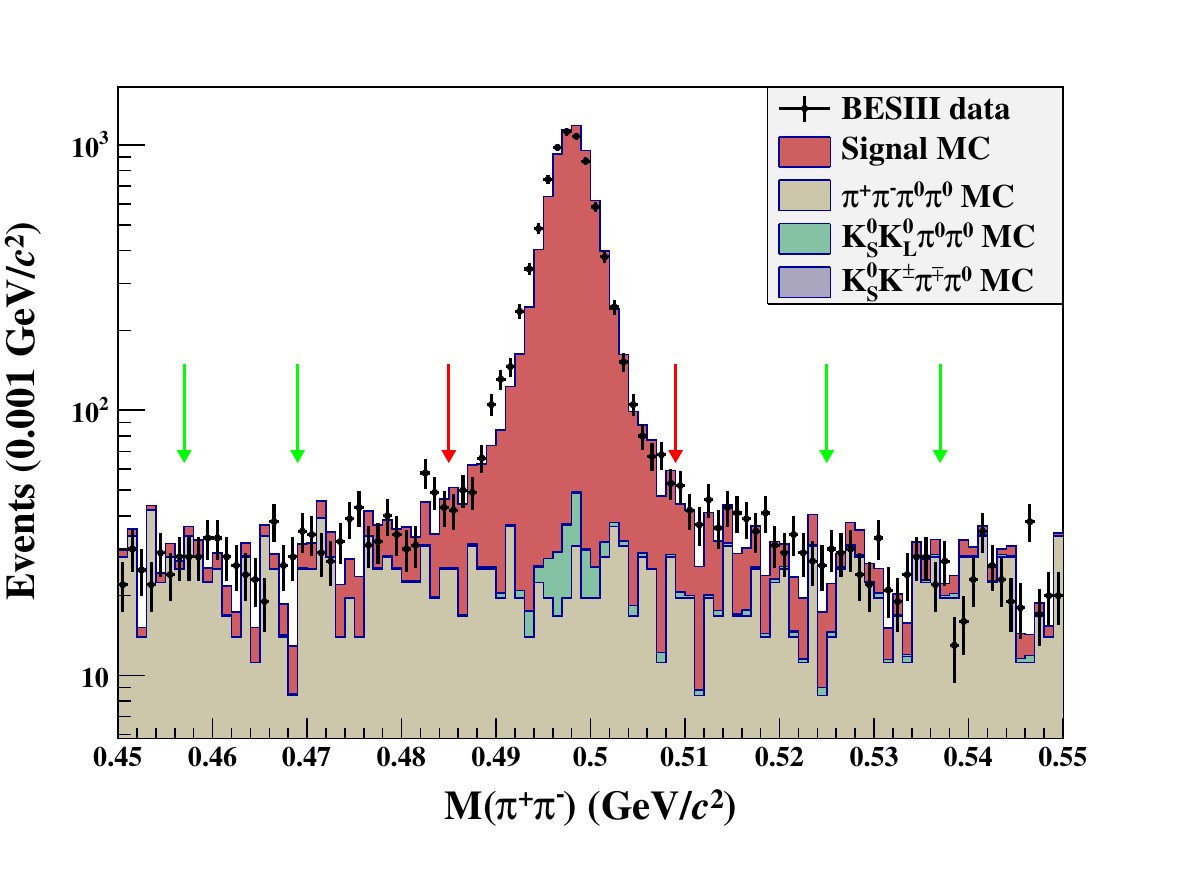}
        \put(15,60){(a)2.125 GeV}
        \end{overpic}
        \begin{overpic}[width=0.48\textwidth]{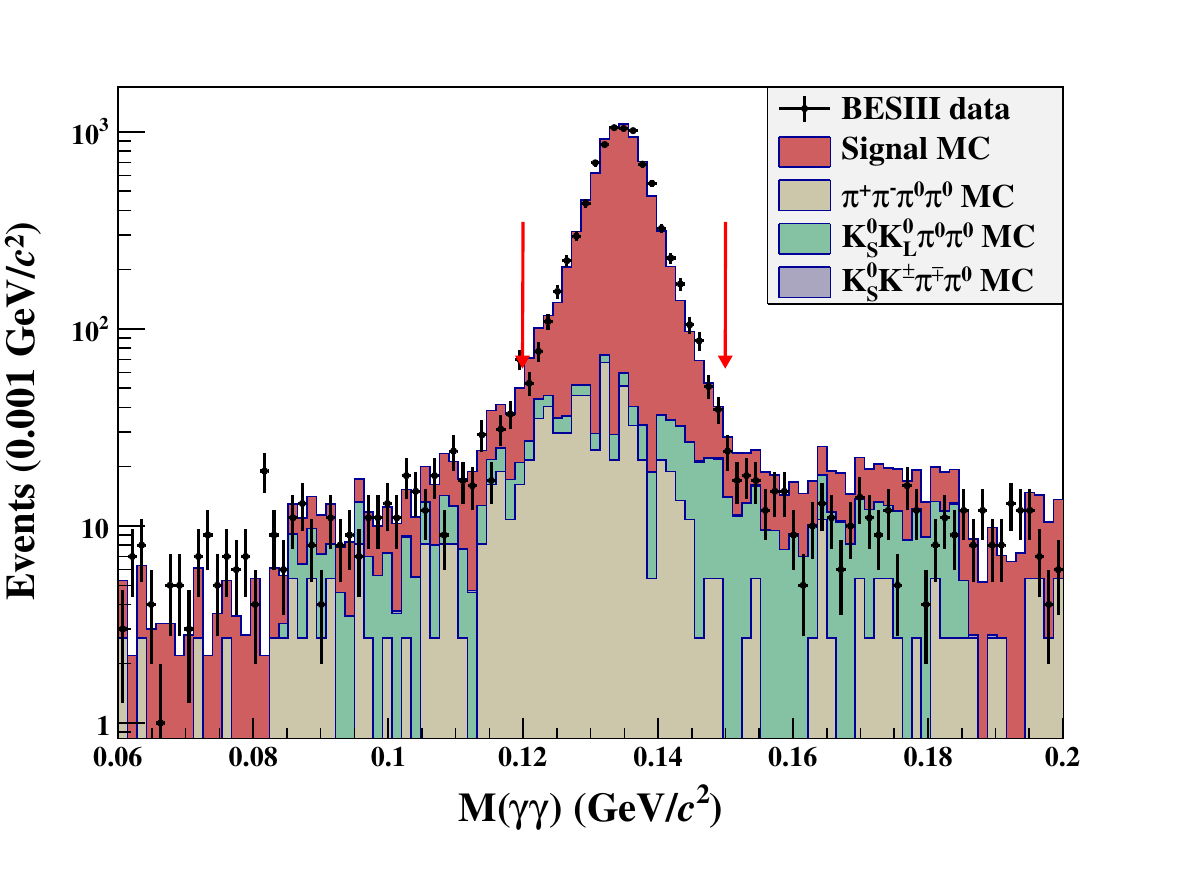}
        \put(15,60){(b)2.125 GeV}
        \end{overpic}
        \begin{overpic}[width=0.48\textwidth]{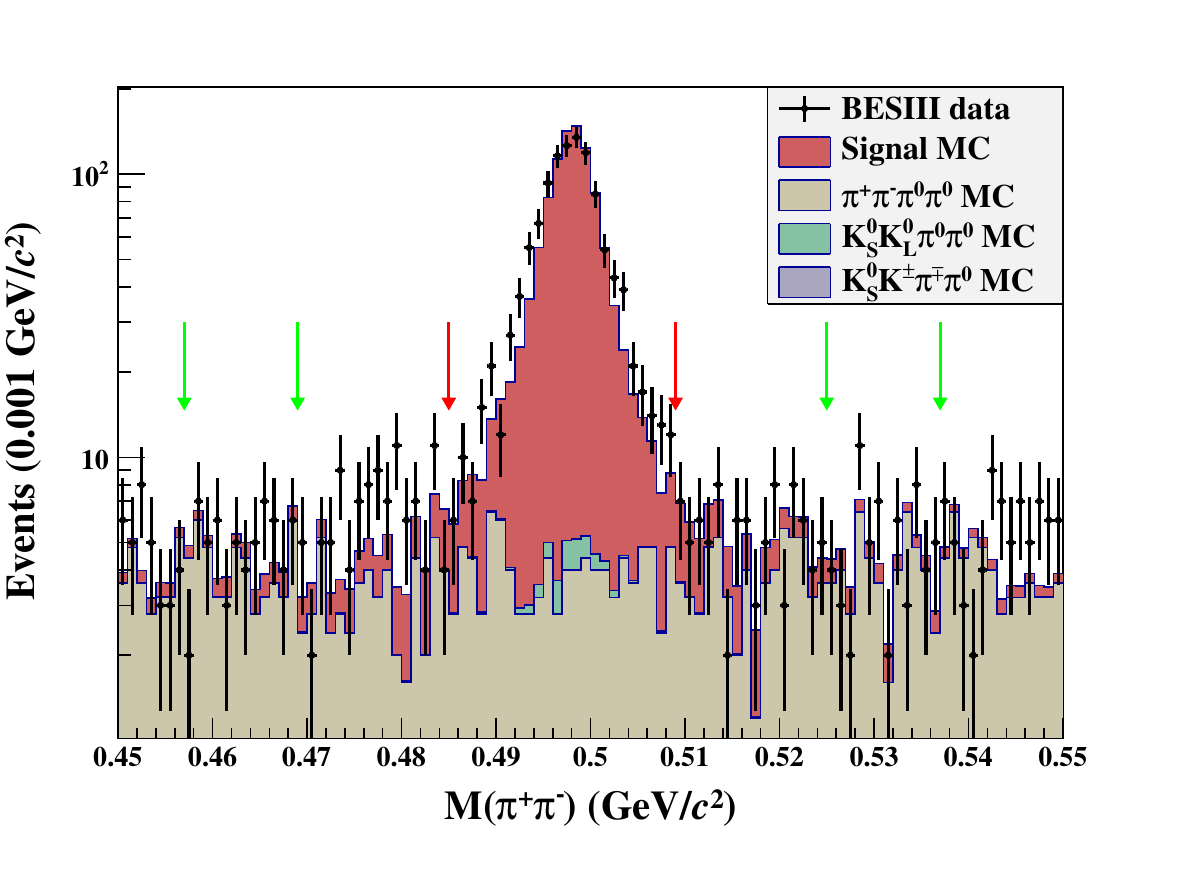}
        \put(15,60){(c)2.900 GeV}
        \end{overpic}
        \begin{overpic}[width=0.48\textwidth]{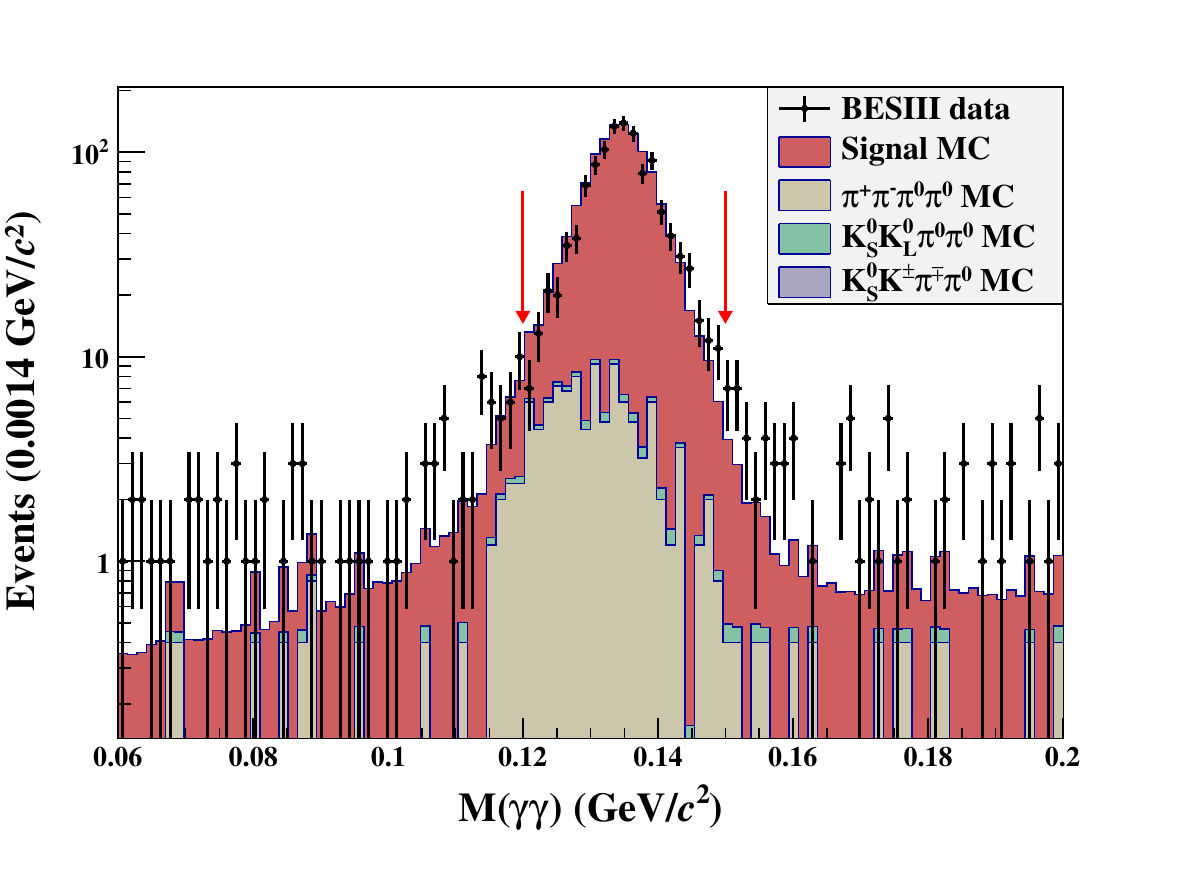}
        \put(15,60){(d)2.900 GeV}
        \end{overpic}
        \flushleft
        \caption{Distributions of (a),(c) $M(\pi^{+}\pi^{-})$ with $\pi^{0}$ mass window requirement and (b),(d)
        $M(\gamma\gamma)$ at $\sqrt{s}=2.125$ and $2.900$ GeV, where the (black)
        dots with error bars are data, and the shaded histogram are the
        stacked MC samples of the signal process,
        $\pi^{+}\pi^{-}\pi^{0}\pi^{0}$, $\Ks\Kl\pi^{0}\pi^{0}$ and $\Ks K^{\pm}\pi^{\mp}\pi^{0}$. The region between red
        arrows is the signal region, and the regions between the green
        arrows are the sideband regions.}
        \label{fig:ksandpi0}
    \end{figure}

Potential background sources are studied by analyzing inclusive
$e^{+}e^{-}\to \rm{hadrons}$ and exclusive $e^{+}e^{-}\to
\pi^{+}\pi^{-}\pi^{0}\pi^{0}$, $\Ks K^{\pm}\pi^{\mp}\pi^{0}$ and
$\Ks\Kl\pi^{0}\pi^{0}$ MC samples after applying the same event
selection criteria. The dominant background process is $e^{+}e^{-}\to
\pi^{+}\pi^{-}\pi^{0}\pi^{0}$. Exclusive
$e^{+}e^{-}\to\pi^{+}\pi^{-}\pi^{0}\pi^{0}$ events are generated by
{\sc{PHOKHARA}}~\cite{bes_phokhara} based on the results of the BaBar
collaboration~\cite{babar_4pi}. The $e^{+}e^{-}\to \Ks
K^{\pm}\pi^{\mp}\pi^{0}$ and $e^{+}e^{-}\to \Ks\Kl\pi^{0}\pi^{0}$ events are generated by {\sc{ConExc}} based
on the dressed cross sections for $e^{+}e^{-}\to \Ks
K^{\pm}\pi^{\mp}\pi^{0}$ and $e^{+}e^{-}\to \Ks\Kl\pi^{0}\pi^{0}$ from the BaBar
experiment~\cite{babar_kskpipi0,babar_ksklpi} with a phase space model and
re-weighted to improve the agreement with BESIII data using a
multidimensional gradient-boosting algorithm~(HEPML)~\cite{hepml}, respectively. The exclusive
$e^{+}e^{-}\to \pi^{+}\pi^{-}\pi^{0}\pi^{0}, \Ks
K^{\pm}\pi^{\mp}\pi^{0}$ and $\Ks\Kl\pi^{0}\pi^{0}$ samples, which have
been normalized to the experimental integrated luminosity, are used to
evaluate the numbers of background events. The contribution of $\Ks$ peaking background events from
$e^{+}e^{-}\to \Ks K^{\pm}\pi^{\mp}\pi^{0}$ and $\Ks\Kl\pi^{0}\pi^{0}$
is at a level of $0.1\%-0.4\%$ for different energy points, which is
negligible in the following fit. The background levels are summarized in table~\ref{background}. Figure~\ref{fig:ksandpi0} shows distributions of the
invariant masses of $\pi^{+}\pi^{-}$, $M(\pi^{+}\pi^{-})$ and
$\gamma\gamma$, $M(\gamma\gamma)$ without the $\Ks$ and $\pi^{0}$ mass window requirements, respectively. Non-$\Ks$ events are characterized
by a flat shape in $M(\pi^{+}\pi^{-})$ and are estimated with the events in the $\Ks$
sideband, which is defined by $0.022
<|M(\pi^{+}\pi^{-})-m_{\Ks}|<0.035~{\rm{GeV}}/c^{2}$.

    \begin{table}[t]
        \centering
        \begin{tabular}{l|c|c}
            \hline
            $\sqrt{s}$~(GeV)    & Background level  & Peaking level \\ \hline
            2.0000              & 5.4\%             & 0.1\%  \\
            2.0500              & 5.2\%             & 0.1\%  \\
            2.1000              & 6.9\%             & 0.1\%  \\
            2.1250              & 6.8\%             & 0.1\%  \\
            2.1500              & 6.4\%             & 0.1\%  \\
            2.1750              & 5.8\%             & 0.1\%  \\
            2.2000              & 6.1\%             & 0.1\%  \\
            2.2324              & 6.3\%             & 0.1\%  \\
            2.3094              & 6.9\%             & 0.1\%  \\
            2.3864              & 6.9\%             & 0.1\%  \\
            2.3960              & 7.1\%             & 0.2\%  \\
            2.6444              & 8.1\%             & 0.2\%  \\
            2.6464              & 8.4\%             & 0.4\%  \\
            2.9000              & 8.1\%             & 0.4\%  \\
            2.9500              & 9.4\%             & 0.4\%  \\
            2.9810              & 8.7\%             & 0.2\%  \\
            3.0000              & 7.2\%             & 0.3\%  \\
            3.0200              & 8.3\%             & 0.2\%  \\
            3.0800              & 8.4\%             & 0.2\%  \\
            \hline
        \end{tabular}
        \caption{Summary of the background level for each $\sqrt{s}$.}
        \label{background}
    \end{table}

\begin{figure}[t]
        \centering
        \begin{overpic}[width=0.48\textwidth]{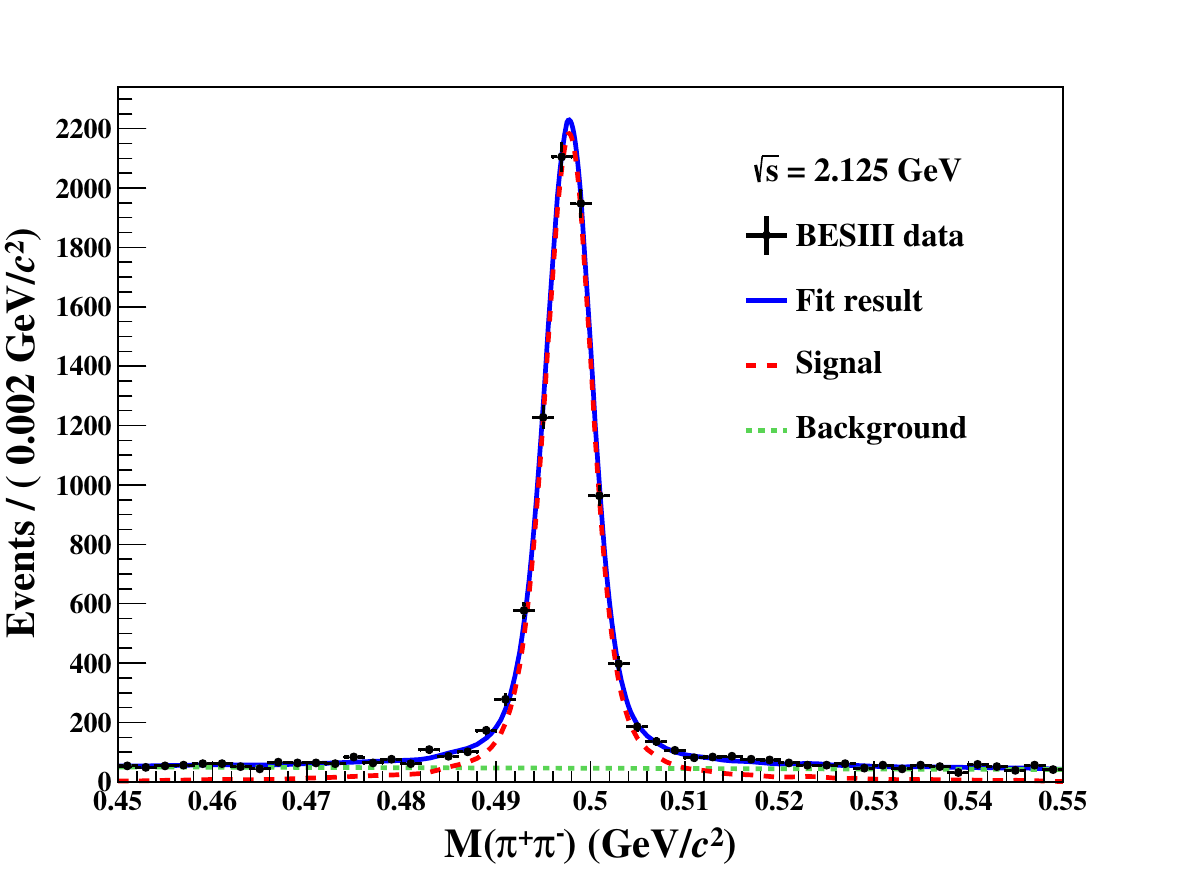}
        \put(15,60){(a)}
        \end{overpic}
        \begin{overpic}[width=0.48\textwidth]{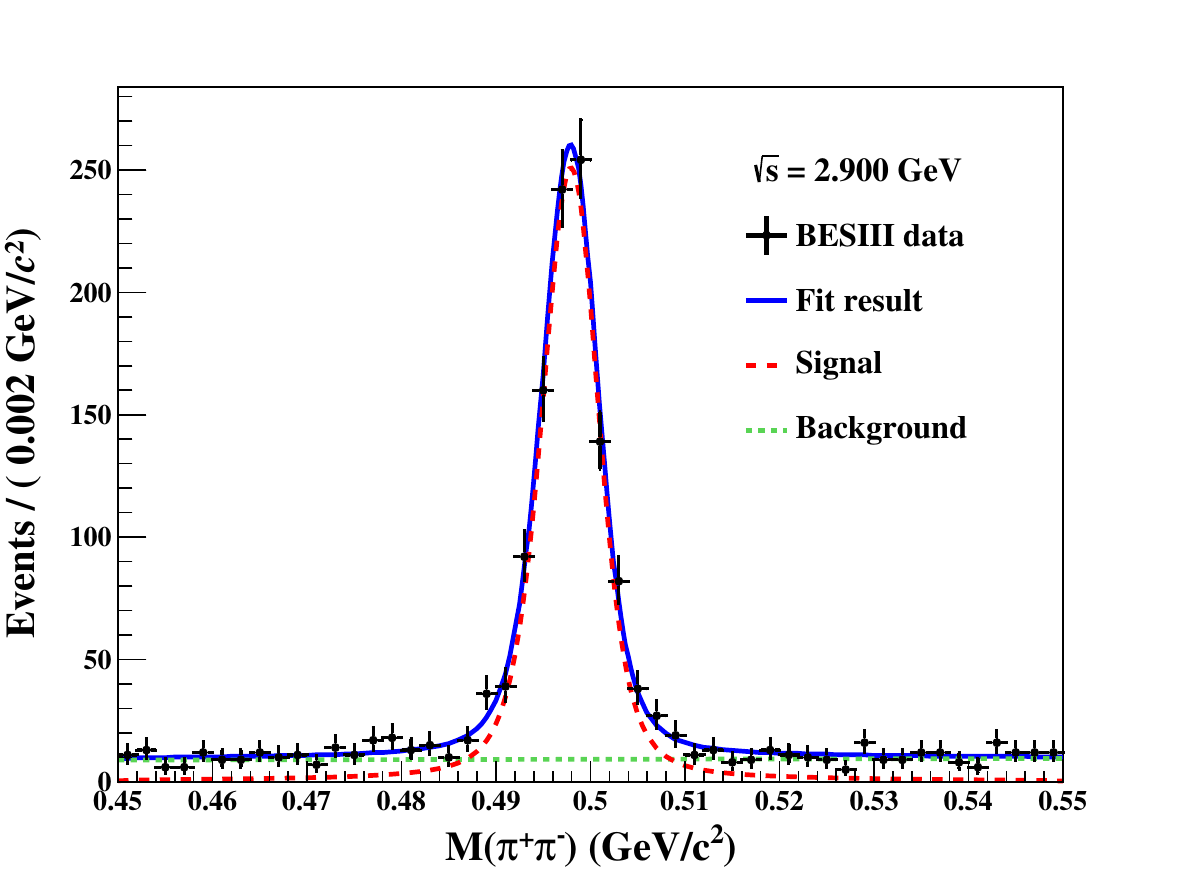}
        \put(15,60){(b)}
        \end{overpic}
        \flushleft
        \caption{Fit to the $M(\pi^{+}\pi^{-})$ distribution at
        $\sqrt{s}=2.125$ and $2.900$ GeV, where the black dots with error bars
        are data, the blue solid curve is the total fit result, the
        green dashed curve indicates the fitted background shape, and
        the red dashed curve is the fitted signal shape.}
        \label{fig:signalfit}
    \end{figure}

The signal yields of the $e^{+}e^{-}\to \Ks\Kl\pi^{0}$ process are
obtained by performing an unbinned maximum likelihood fit to the
$M(\pi^{+}\pi^{-})$ spectrum. The signal component is described by the signal
MC-simulated shape convolved with a Gaussian function which describes
the difference between data and MC simulation. The mean value and width of the Gaussian function are separately floated parameters at different energy points. The background function
is parameterized by a first-order polynomial function. The
corresponding fit results for data taken at \mbox{$\sqrt{s}=2.125$} and \mbox{$2.900$ GeV}
are shown in figure~\ref{fig:signalfit}. The same event selection
criteria and fit procedure are applied for all data samples at the
nineteen c.m. energies.

\begin{figure}[t]
        \centering
        \begin{overpic}[width=0.45\textwidth]{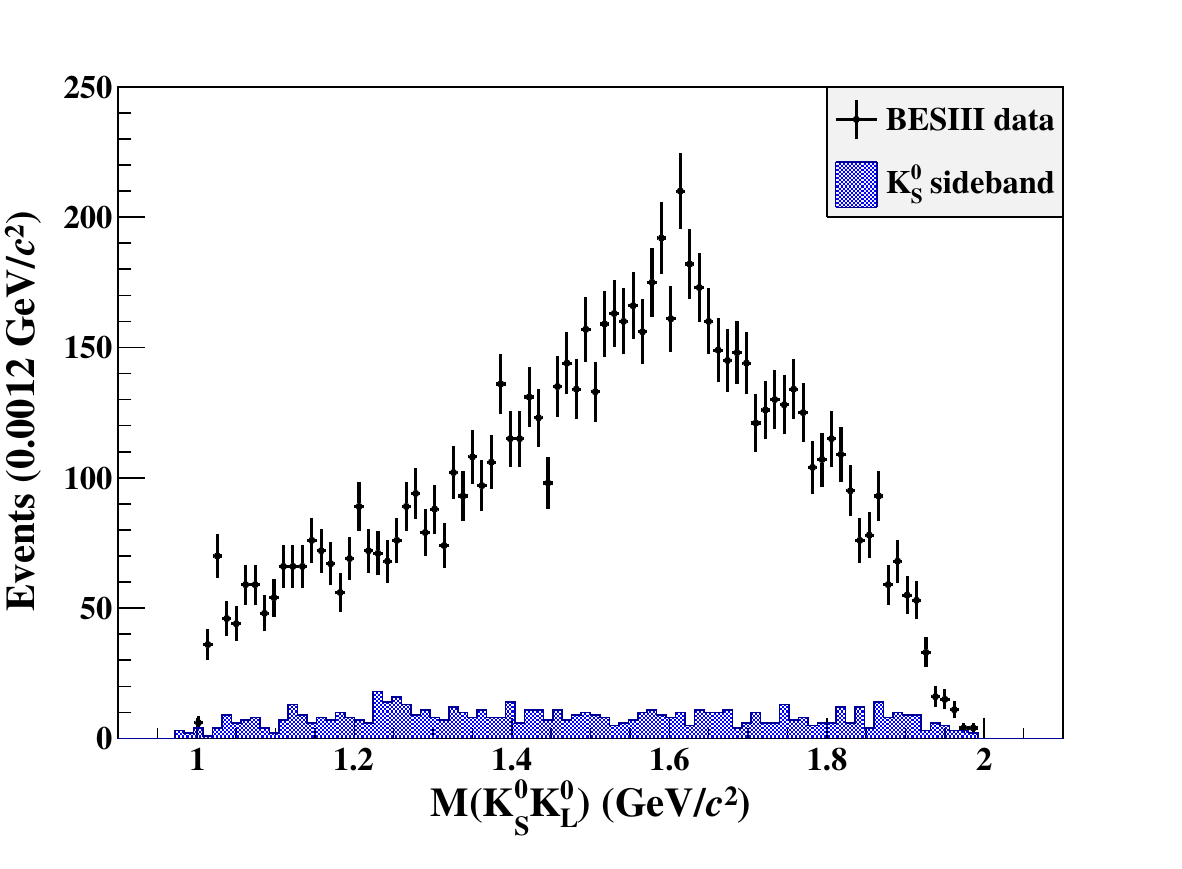}
        \put(15,55){(a)}
        \end{overpic}
        \begin{overpic}[width=0.45\textwidth]{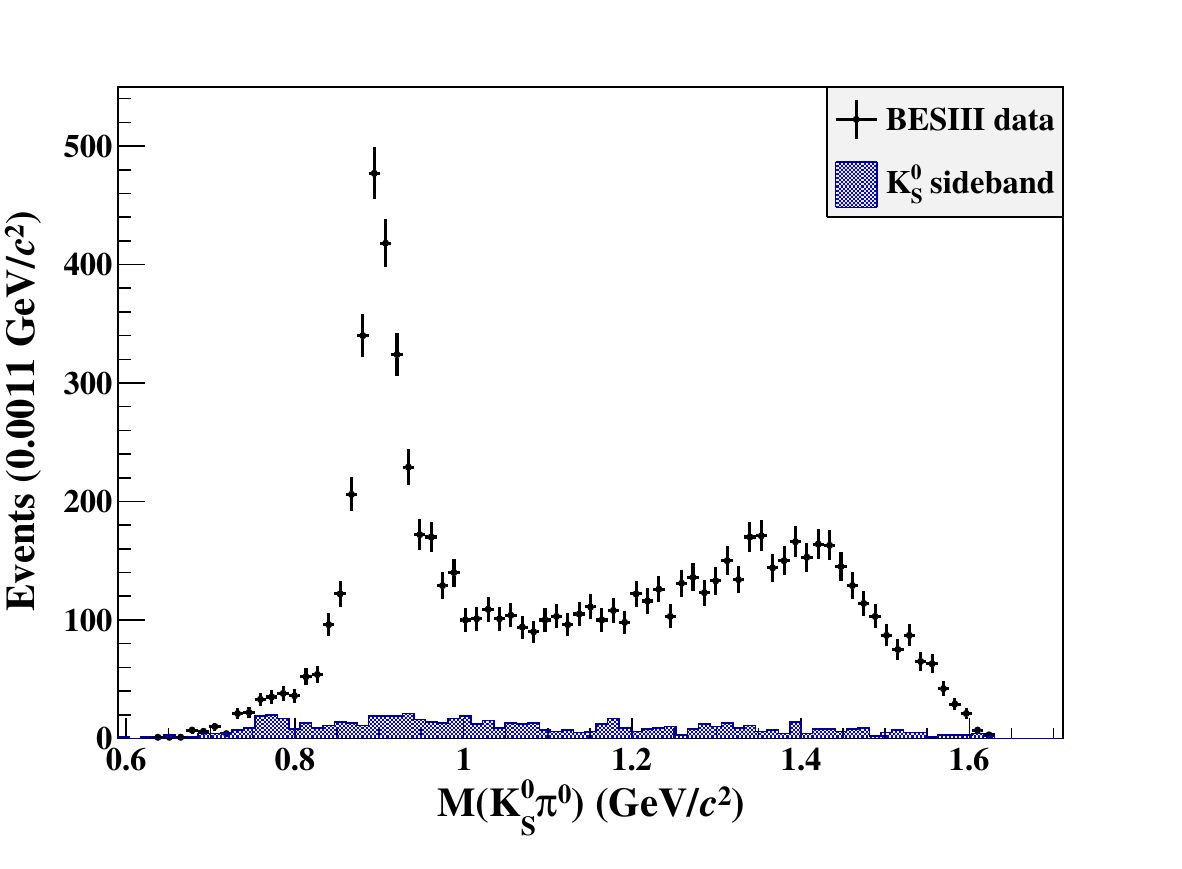}
        \put(15,55){(b)}
        \end{overpic}
        \begin{overpic}[width=0.45\textwidth]{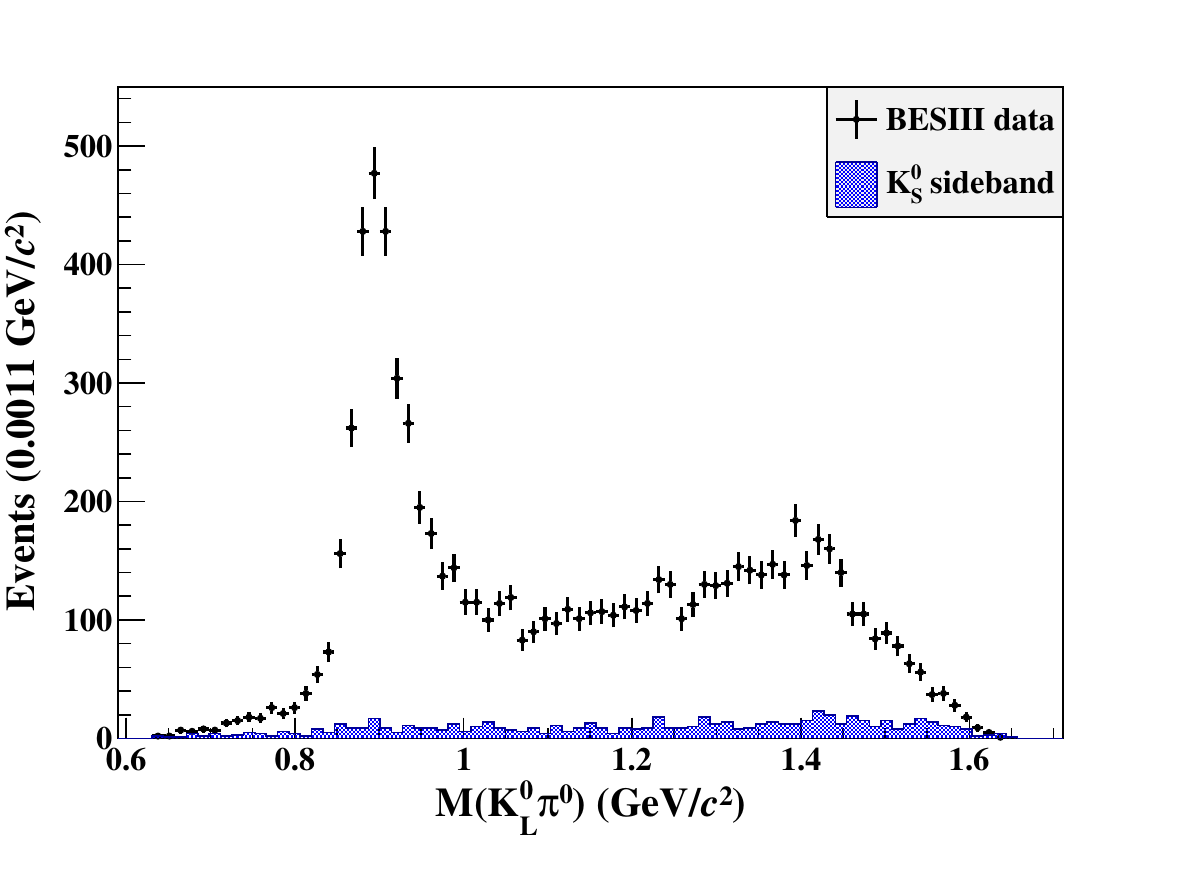}
        \put(15,55){(c)}
        \end{overpic}
        \begin{overpic}[width=0.45\textwidth]{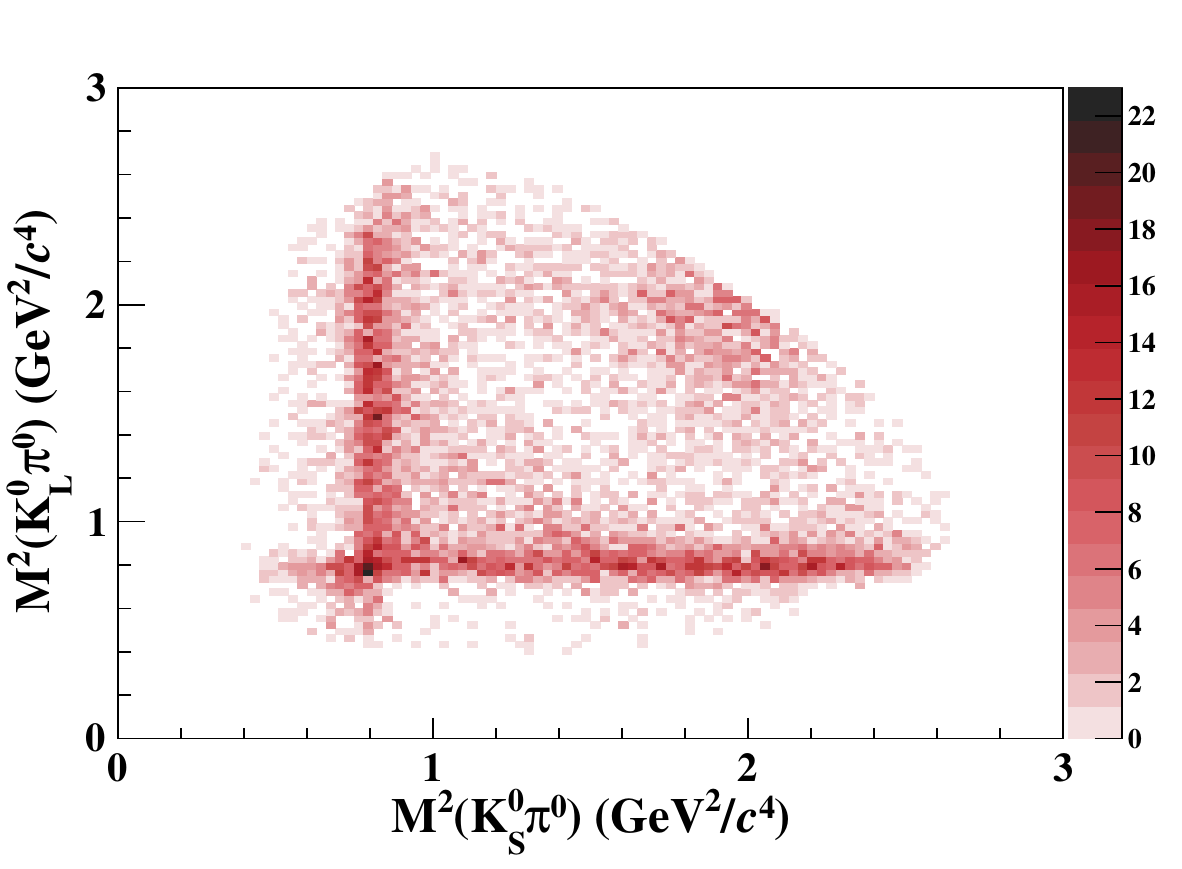}
        \put(15,55){(d)}
        \end{overpic}
        \flushleft
        \caption{Distributions of (a) $M(\Ks\Kl)$, (b) $M(\Ks\pi^{0})$
        and (c) $M(\Kl\pi^{0})$, where the (black) dots with error
        bars are data, and the shaded histograms are non-$\Ks$ events
        estimated by the $\Ks$ sideband. (b) Distribution of
        $M^{2}(\Ks\pi^{0})$ versus $M^{2}(\Kl\pi^{0})$. All plots are
        based on data at $\sqrt{s}=2.125$ GeV.}
        \label{fig:2body_ori}
    \end{figure}

In order to improve the resolution of kinematic variables, the
remaining \mbox{$e^{+}e^{-}\to\Ks\Kl\pi^{0}$} events are subjected to
a three-constraint~(3C) kinematic fit, which, in addition to imposing
energy and momentum conservation, further constrains the $\pi^{0}$ and
$\Ks$ masses to their PDG values~\cite{PDG}. After
all above criteria, the invariant mass spectra of
$\Ks\Kl$, $\Ks\pi^{0}$, $\Kl\pi^{0}$ and the invariant masses squared of
$\Ks\pi^{0}$ versus $\Kl\pi^{0}$ are shown in
figure~\ref{fig:2body_ori}, where the $\Kstar$ structure is clear. For the
invariant mass spectra of $\Ks\Kl$, $\Ks\pi^{0}$ and $\Kl\pi^{0}$, the
contributions of background events which are obtained by the $\Ks$
sideband are smooth and
confirm that there is no peaking structure. Those non-$\Ks$ events are
used to estimate the background contributions and those $\Ks$ peaking backgrounds are negligible in the following
amplitude analysis.

\section{Amplitude analysis} 

Based on the GPUPWA framework~\cite{gpupwa}, a PWA is performed on the
surviving candidate events to identify the intermediate processes
present in $e^{+}e^{-}\to\Ks\Kl\pi^{0}$. The quasi-two-body decay
amplitudes in the process $e^{+}e^{-}\to\Ks\Kl\pi^{0}$ with two
sequential decays \mbox{$e^{+}e^{-}\to R_{1}\pi^{0}\to\Ks\Kl\pi^{0}$
and $e^{+}e^{-}\to \Ks(\Kl) R_{2}\to\Ks\Kl\pi^{0}$} are considered and
constructed using the covariant tensor amplitude
formalism~\cite{zoubs}, where $R_{1}$ and $R_{2}$ are the intermediate
states that can decay to $\Ks\Kl$ and $\Ks(\Kl)\pi^{0}$, respectively.

According to Ref.~\cite{zoubs}, the general form for the decay
amplitude of a $1^{-}$ state ($Y$) is \begin{equation}
A(m) = Y_{\mu}(m)A^{\mu}=Y_{\mu}(m)\sum_{i}\Lambda_{i}U_{i}^{\mu},
\end{equation} 

\noindent where $Y_{\mu}(m)$ is the polarization vector of $Y$, $m$ is the
spin projection of $Y$, and $U_{i}^{\mu}$ is the \mbox{$i$-th}
partial-wave amplitude with coupling strength determined by a complex
parameter $\Lambda_{i}$. The amplitude $U^{\mu}_{i}$ is constructed with the spin factor, Blatt-Weisskopf barrier factors and propagators of resonances under the assumption of isobar model~\cite{zoubs}. The differential cross-section can be written as
    \begin{equation}
        \frac{d\sigma}{d\Phi_{n}}=\frac12\sum_{m}A^{\mu}A^{*\mu}=
        \frac12\sum_{i,j}\Lambda_{i}\Lambda_{j}^{*}\sum_{m}U^{\mu}_{i}U^{*\mu}_{j}.
    \end{equation}

The spin factor is constructed with the covariant Zemach~(Rarita-Schwinger) tensor formalism~\cite{zoubs,sf1,sf2,sf3} by combining pure-orbital-angular-momentum covariant tensors $\tilde{t}_{\mu_1...\mu_L}^{(L)}$ and 
the momenta of parent particles together with Minkowski metric $g_{\mu\nu}$ and Levi-Civita symbol $\epsilon_{\mu\nu\lambda\sigma}$. For a process $a\to bc$, the covariant tensors $\tilde{t}_{\mu_1...\mu_L}^{(L)}$ for the final states of pure orbital angular momentum $L$ are
    \begin{equation}
        \tilde{t}^{(L)}_{\mu_1...\mu_L}=(-1)^{L}P^{(L)}_{\mu_1...\mu_L\mu_1^{'}...\mu_L^{'}}(p_a)r^{\mu_1^{'}...\mu_L^{'}B_{L}(Q_{abc})},
    \end{equation}
where $r=p_b-p_c$, $P^{(L)}_{\mu_1...\mu_L\mu_1^{'}...\mu_L^{'}}(p_a)$ is the spin projection operator of the particle $a$, $Q_{abc}$ is the magnitude of $p_b$ or $p_c$ in the rest system of $a$. The Blatt-Weisskopf barrier factors $B_{L}(Q_{abc})$, are derived by assuming a square well interaction potential as 
    \begin{align}
        B_{0}(Q_{abc})&=1, \\
        B_{1}(Q_{abc})&=\sqrt{\frac{2}{Q^{2}_{abc}+Q_{0}^{2}}}, \\
        B_{2}(Q_{abc})&=\sqrt{\frac{13}{Q^{4}_{abc}+3Q^{2}_{abc}Q_{0}^{2}+9Q^{4}_{0}}}.
    \end{align}
Here $Q^{0}=0.197321/R$ GeV/$c$ is a hadron "scale" parameter, where $R$ is the radius of the centrifugal barrier in fm. In this paper, the radius $R$ is taken to be 0.7 fm.

The propagator of intermediate resonance is parameterized by a
relativistic Breit-Wigner~(BW) function with an invariant mass
dependent width~\cite{rbw}
\begin{align}
{\rm BW}(s)&=\frac{1}{m^{2}-s-i\sqrt{s}\Gamma(s)},\\
\Gamma(s)&=\Gamma_{0}(m^{2})(\frac{m^{2}}{s})(\frac{p(s)}{p(m^{2})})^{2l+1},
\end{align} 

\noindent where $s$ is the invariant mass squared of the daughter particle, $m$
and $\Gamma_{0}$ are the mass and width of the intermediate resonance,
respectively, $l$ is the orbital angular momentum for a daughter
particle, and $p(s)$ or $p(m^{2})$ is the momentum of a daughter
particle in the rest frame of the resonance with mass $\sqrt{s}$ or
$m$. To include the resolution effect for the narrow $\phi$ resonance,
the BW function is convolved with a Gaussian function.

The relative magnitudes and phases of the individual intermediate
processes are determined by performing an unbinned maximum likelihood
fit using {\sc{MINUIT}}~\cite{minuit}, where the magnitude and phase
of the reference amplitude $e^{+}e^{-}\to\Kstar\Kbar$ are fixed to 1
and 0, respectively, while those of other amplitudes are free 
parameters of the fit. 

The negative log-likelihood function for observing N events in the data sample is expressed as
\begin{equation}
\mathcal{NLL}=-\sum_{i}^{N}\log\frac{\omega_{i}\epsilon_{i}}{\int\epsilon\omega d\Phi_{3}}=-\sum_{i}^{N}\log\frac{\omega_{i}}{\int\epsilon\omega d\Phi_{3}}+const,
\end{equation}
where $\omega_{i}$ is the decay-amplitude squared evaluated from the four-momenta of final particles for the $i$-th event, $\epsilon_{i}$ is the detection efficiency and $\Phi$ is the standard element of phase space. The contribution of background events to the $\mathcal{NLL}$ is canceled out by evaluating the signal model on $\Ks$ sideband events injected into the data sample with negative weights.

Conservation of $J^{PC}$ for intermediate states, in the process
$e^{+}e^{-}\to R_{1}\pi^{0}\to\Ks\Kl\pi^{0}$, allows both
$\mathcal{P}$ and $\mathcal{F}$ wave contributions both in
$e^{+}e^{-}\to R_{1}\pi^{0}$ and $R_{1}\to \Ks\Kl$. In the case of
$e^{+}e^{-}\to \Ks(\Kl) R_{2}\to\Ks\Kl\pi^{0}$, the contributions of
$\mathcal{P}$, $\mathcal{D}$ and $\mathcal{F}$ waves are all allowed
both in the primary and secondary processes. The PWA fit procedure
starts by including the $\Kstar\Kbar$ and $\KstarT\Kbar$ as the
initial baseline solutions, and then adds one at a time other possible
intermediate states which can decay to $\Ks(\Kl)\pi^{0}$ or $\Ks\Kl$.
The masses and widths of possible intermediate resonances are fixed to
their PDG values~\cite{PDG}. Intermediate states are included in the
solution if the statistical significance is greater than 5$\sigma$, where
the statistical significance is evaluated from the changes in
likelihood and degrees of freedom with and without the corresponding
amplitude included in the PWA fit. The direct decay process without an
intermediate resonance is treated as a phase space
distribution without a propagator~\cite{besiii_kkpi}. The procedure is repeated until a
best solution is obtained.

The above strategy is implemented individually on the experimental
data sets collected at \mbox{$\sqrt{s}=2.125,2.396$ and $2.900$ GeV},
which have the largest luminosities and yields among the nineteen data
sets. As the c.m. energy increases, we test the significance of the process with higher threshold at $\sqrt{s}=2.900$ GeV, such as $e^{+}e^{-}\to K_{3}^{*}(1780)^{0}\bar{K}^{0}$, which cannot be produced at $\sqrt{s}=2.125$ GeV. But the significances of these processes are less than 5$\sigma$ and they are not retained in the final best solution. The statistical significances of the intermediate states and fit fractions for \mbox{$\sqrt{s}=2.125$, $2.396$ and $2.900$
GeV} are listed in table~\ref{significance} and
table~\ref{fraction}, respectively. For the other sixteen data samples with lower luminosities and limited statistics, the intermediate components are assumed to be the same as those of the nearby c.m. energies with higher statistics. The intermediate component candidates of $\sqrt{s}=2.000,2.050,2.100,2.150,2.175,2.200$, and $2.232$ GeV are assumed to be the same as $\sqrt{s}=2.125$ GeV. The intermediate component candidates of $\sqrt{s}=2.309,2.386,2.644$ and $2.646$ GeV are assumed to be the same as $\sqrt{s}=2.396$ GeV. The remaining datasets are assumed to use the same intermediate components as $\sqrt{s}=2.900$ GeV. The data for each energy point is fitted individually.

    \begin{table}[t]
        \centering
        \begin{tabular}{l|ccc}
            \hline
            \multirow{2}{*}{Process} &\multicolumn{3}{c}{Significance}\\ \cline{2-4}
                          &2.125 GeV &2.396 GeV &2.900 GeV\\ \hline
            $\phi\pi^{0}$ & 13.1$\sigma$ & 8.6$\sigma$ & 9.7$\sigma$\\
            $\phi(1680)\pi^{0}$ & 11.1$\sigma$ & 12.2$\sigma$ & 8.3$\sigma$\\
            $\Kstar\Kbar$ & >30$\sigma$ & >30$\sigma$ & >30$\sigma$\\
            $\KstarT\Kbar$ & 29.2$\sigma$ & 5.7$\sigma$ & 5.1$\sigma$\\
            $K(1680)^{0}\Kbar$ & 9.8$\sigma$ & 8.4$\sigma$ & 7.6$\sigma$\\
            \hline
        \end{tabular}
        \caption{Statistical significances of the intermediate states for data at $\sqrt{s}=2.125,2.396$ and $2.900$ GeV.}
        \label{significance}
    \end{table}

    \begin{table}[t]
        \centering
        \begin{tabular}{l|ccc}
            \hline
            \multirow{2}{*}{Process} &\multicolumn{3}{c}{Fraction~(\%)}\\ \cline{2-4}
                          &2.125 GeV &2.396 GeV &2.900 GeV\\ \hline
            $\phi\pi^{0}$       & 0.78$\pm$0.56     & 0.87$\pm$0.61     & 1.82$\pm$1.11\\
            $\phi(1680)\pi^{0}$ & 2.39$\pm$1.23     & 5.96$\pm$2.10     & 5.22$\pm$1.50\\
            $\Kstar\Kbar$       & 79.89$\pm$1.12    & 86.01$\pm$1.38    & 72.65$\pm$2.11\\
            $\KstarT\Kbar$      & 7.42$\pm$0.83     & 1.93$\pm$0.57     & 1.85$\pm$0.82\\
            $K(1680)^{0}\Kbar$  & 3.00$\pm$1.11     & 6.73$\pm$1.91     & 5.82$\pm$1.96\\
            \hline
        \end{tabular}
        \caption{Fit fractions of the intermediate states for data at $\sqrt{s}=2.125,2.396$ and $2.900$ GeV.}
        \label{fraction}
    \end{table}

\begin{figure}[t]
        \centering
        \begin{overpic}[width=0.48\textwidth]{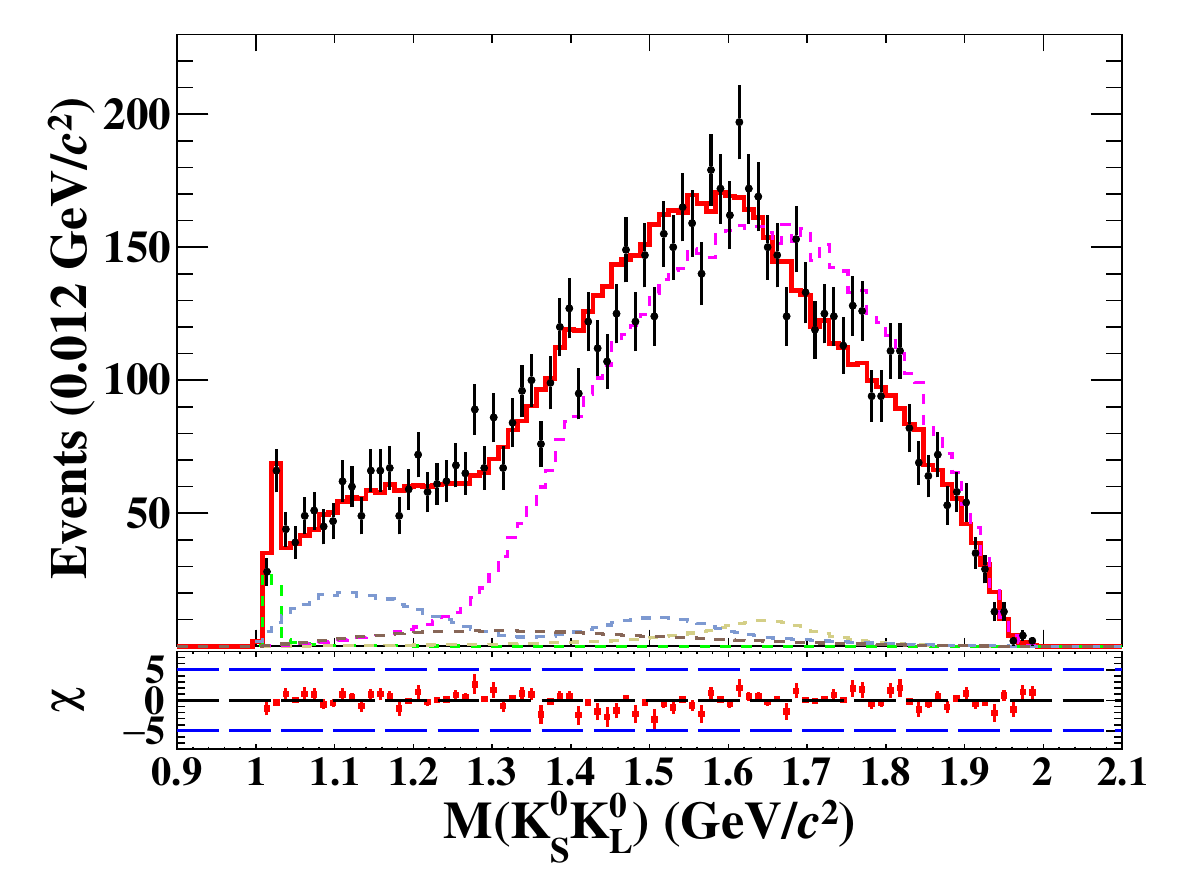}
        \put(17,65){(a)}
        \end{overpic}
        \begin{overpic}[width=0.48\textwidth]{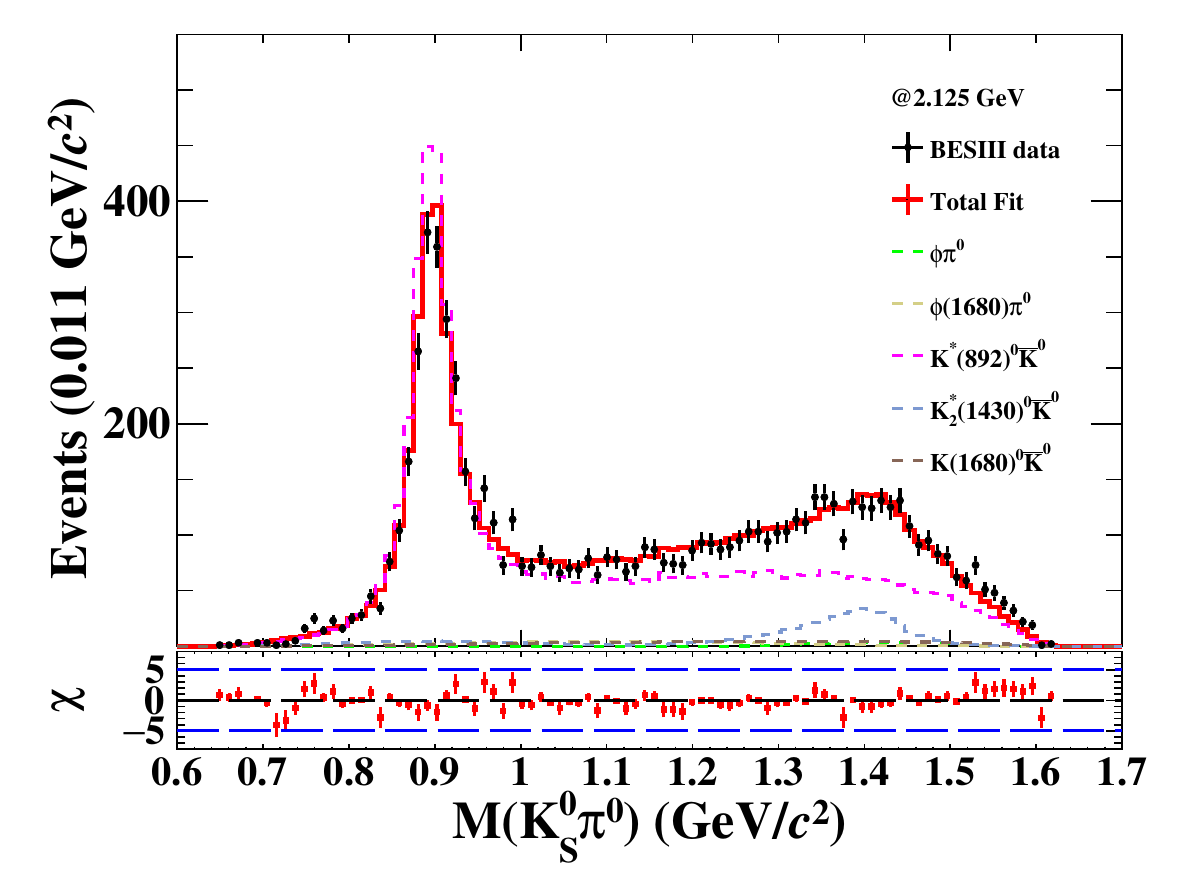}
        \put(17,65){(b)}
        \end{overpic}
        \begin{overpic}[width=0.48\textwidth]{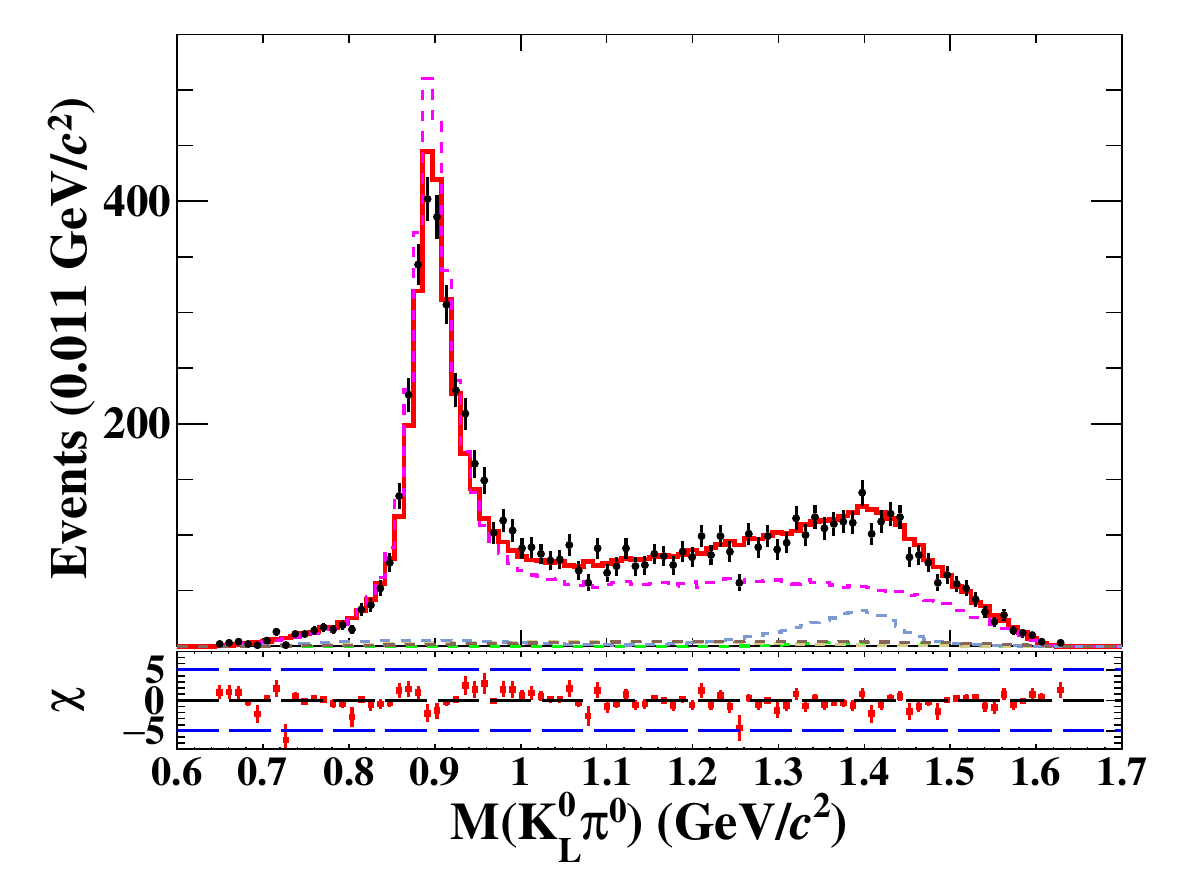}
        \put(17,65){(c)}
        \end{overpic}
        \begin{overpic}[width=0.48\textwidth]{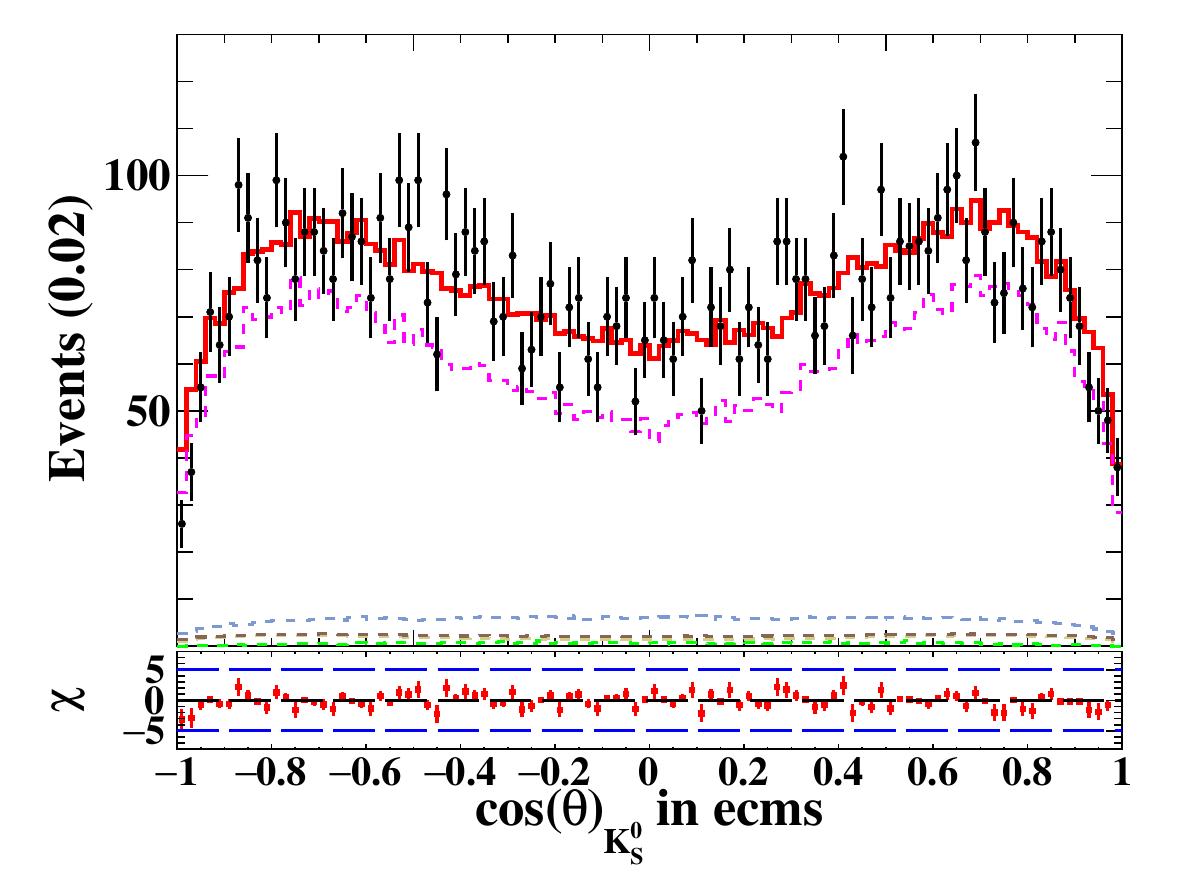}
        \put(17,65){(d)}
        \end{overpic}
        \begin{overpic}[width=0.48\textwidth]{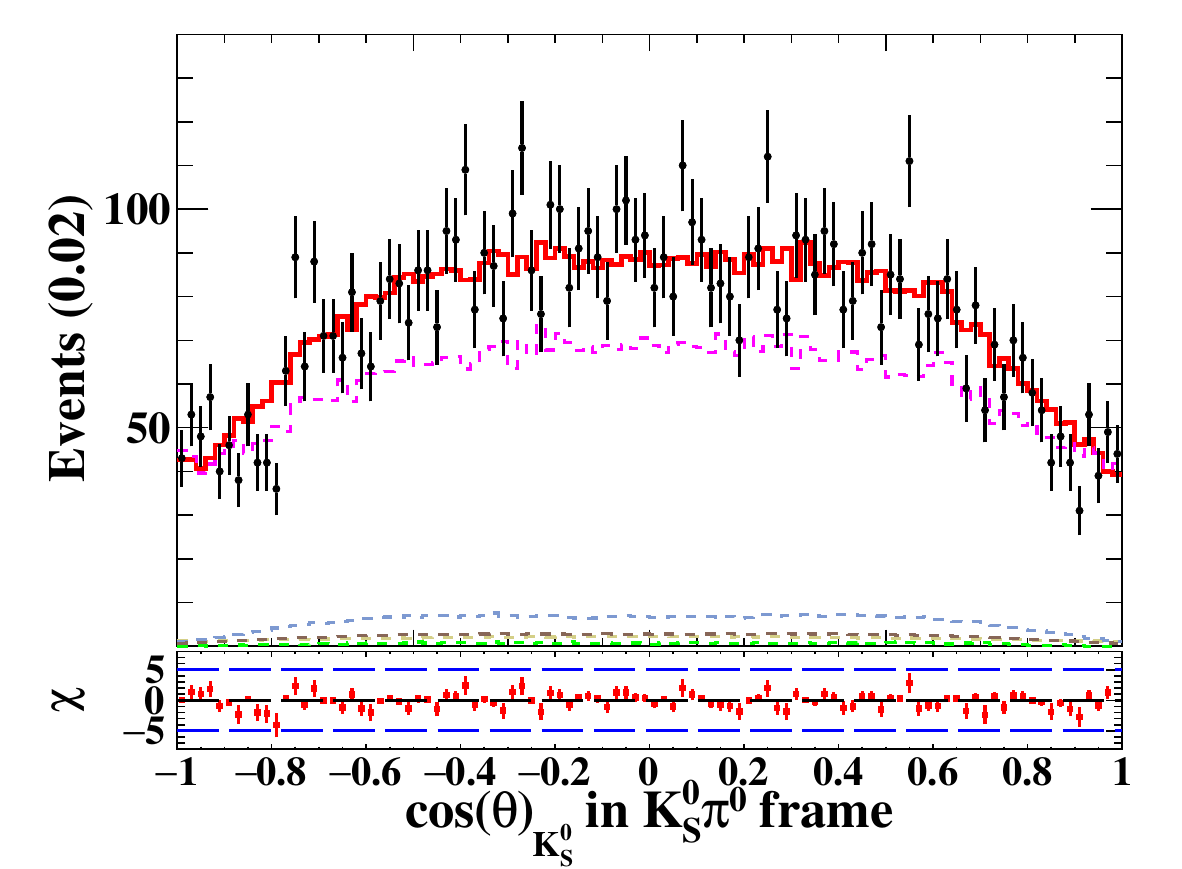}
        \put(17,65){(e)}
        \end{overpic}
        \begin{overpic}[width=0.48\textwidth]{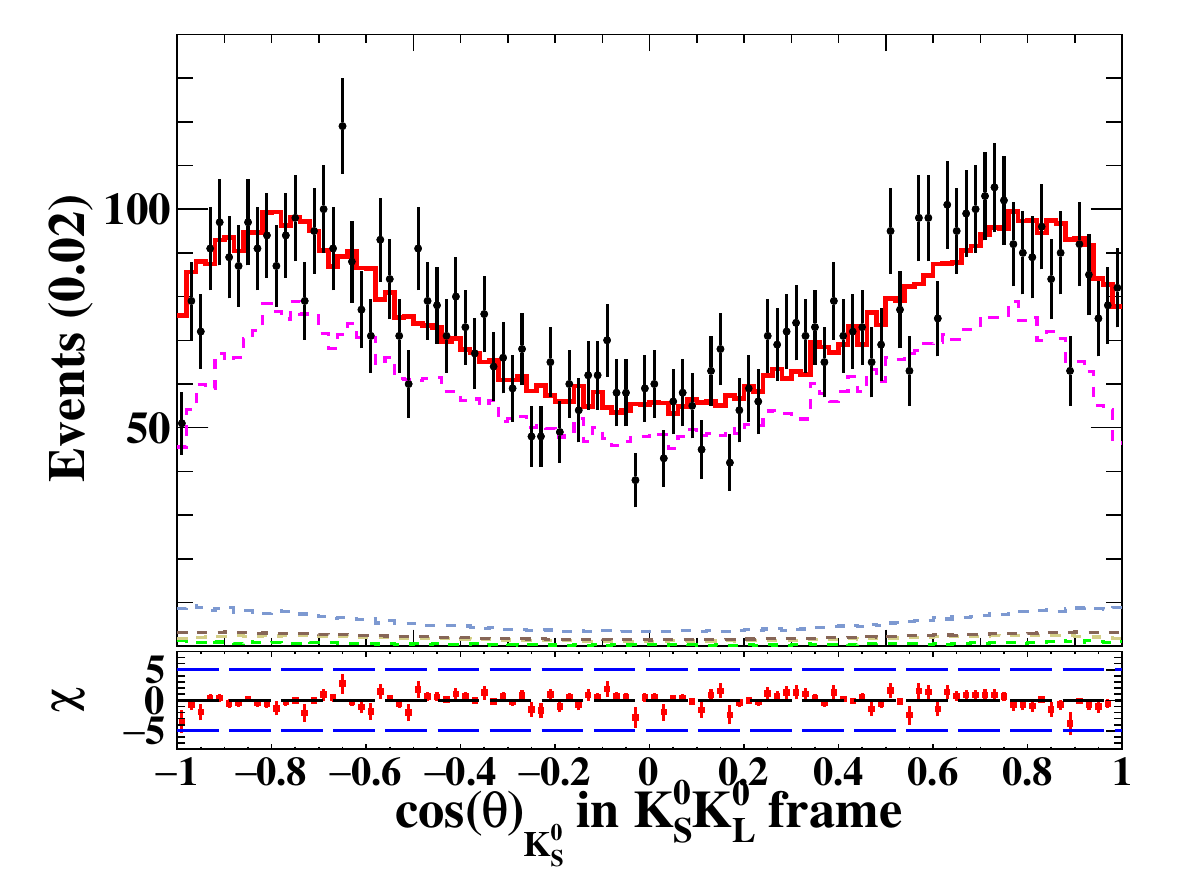}
        \put(17,65){(f)}
        \end{overpic}
        \flushleft
        \caption{Superposition of data and the PWA fit projections for
        invariant mass distributions of (a) $\Ks\Kl$, (b) $\Ks\pi^{0}$
        and (c) $\Kl\pi^{0}$, and the $\!\cos\theta$ distributions of (d)
        $\Ks$ in $e^{+}e^{-}$ c.m. frame, (e) $\Ks$ in $\Ks\pi^{0}$
        rest frame and (f) $\Ks$ in $\Ks\Kl$ rest frame at
        $\sqrt{s}=2.125$ GeV. The pull projection of the residuals
        is shown beneath each distribution correspondingly. Different
        styles of the curves denote different components.}
        \label{fig:2125pwa}
    \end{figure}

The invariant mass spectra, angular distributions and fit results
for \mbox{$\sqrt{s}=2.125$ GeV} are shown in
figure~\ref{fig:2125pwa}. 

\section{Born cross sections measurement} 

The Born cross section for $e^{+}e^{-}\to \Ks\Kl\pi^{0}$ is obtained
at each c.m. energy using \begin{equation} \label{cs}
\sigma(\sqrt{s})=\frac{N_{\rm{sig}}}{\mathcal{L}\cdot\epsilon\cdot(1+\delta)\cdot\frac{1}{|1-\Pi|^{2}}\cdot\mathcal{B}},
\end{equation} where $N_{\rm{sig}}$ is the number of signal events,
$\mathcal{L}$ is the integrated luminosity, $\epsilon$ is the
efficiency obtained by weighting MC simulation according to the PWA
results, $\mathcal{B}$ is the product of BFs in the full decay chain
$\mathcal{B}=\mathcal{B}(\Ks\to\pi^{+}\pi^{-})\cdot\mathcal{B}(\pi^{0}\to\gamma\gamma)=68.39\%$,
which is taken from the PDG~\cite{PDG}, $\frac{1}{|1-\Pi|^{2}}$ is the
vacuum polarization~(VP) factor~\cite{VP}, and $1+\delta$ is the ISR
correction factor, which is obtained by a QED
calculation~\cite{isr_calculate}. Both $\epsilon$ and $1+\delta$
depend on the line shape of cross sections and are determined by an
iterative procedure~\cite{jingmq,bes_kskl}.  The Born cross section
for an intermediate process, $e^{+}e^{-}\to \Kstar\Kbar$ or
$\KstarT\Kbar$, at each energy is obtained with the same approach,
where $N_{\rm{sig}}$ is replaced with the product of the total number
of surviving events and the corresponding fraction relative to the
total obtained according to the PWA results, and $\mathcal{B}$ is
replaced with the product of the BFs of the decays
$\Ks\to\pi^{+}\pi^{-},\pi^{0}\to\gamma\gamma$ and that of the
intermediate state~($\Kstar\to K^{0}\pi^{0}=33.23\%,\KstarT\to K^{0}\pi^{0}=16.60\%$) from the PDG~\cite{PDG}, respectively. The Born cross sections
are listed in tables~\ref{tab:cs3}, \ref{tab:csk892}
and~\ref{tab:csk2}, separately for the processes
\mbox{$e^{+}e^{-}\to\Ks\Kl\pi^{0}$, $\Kstar\Kbar$ and $\KstarT\Kbar$},
respectively.

The previous BESIII measurement \cite{besiii_kkpi} with the charged channel
$e^{+}e^{-}\to K^{+}K^{-}\pi^{0}$ shows that
$\KstarcT K^{-}$ is the dominant component, with the fraction of
$\Kstarc K^{-}$ at the 2-10\% level. However, in this study with the neutral
channel $e^{+}e^{-}\to\Ks\Kl\pi^{0}$,
$\Kstar\Kbar$ is dominant, while $\KstarT\Kbar$ is at the 5\% level in the
BESIII c.m. energy region. The asymmetry is also
observed by BaBar~\cite{babar_kskpi} in the production of $\Kstar\Kbar$, $\Kstarc K^{-}$,
$\KstarT\Kbar$ and $\KstarcT K^{-}$. To quantify the effect, we define
relative ratios of the Born cross sections:
    \begin{equation}
        \begin{aligned}
            {\it{R}}(K^{*}(892))&=\frac{\sigma(e^{+}e^{-}\to\Kstarc K^{-})}{\sigma(e^{+}e^{-}\to\Kstar\Kbar)},\\
            {\it{R}}(K^{*}_{2}(1430))&=\frac{\sigma(e^{+}e^{-}\to\KstarcT K^{-})}{\sigma(e^{+}e^{-}\to\KstarT\Kbar)}.
        \end{aligned}
    \end{equation}

\begin{figure}[t]
        \centering
        \begin{overpic}[width=0.48\textwidth]{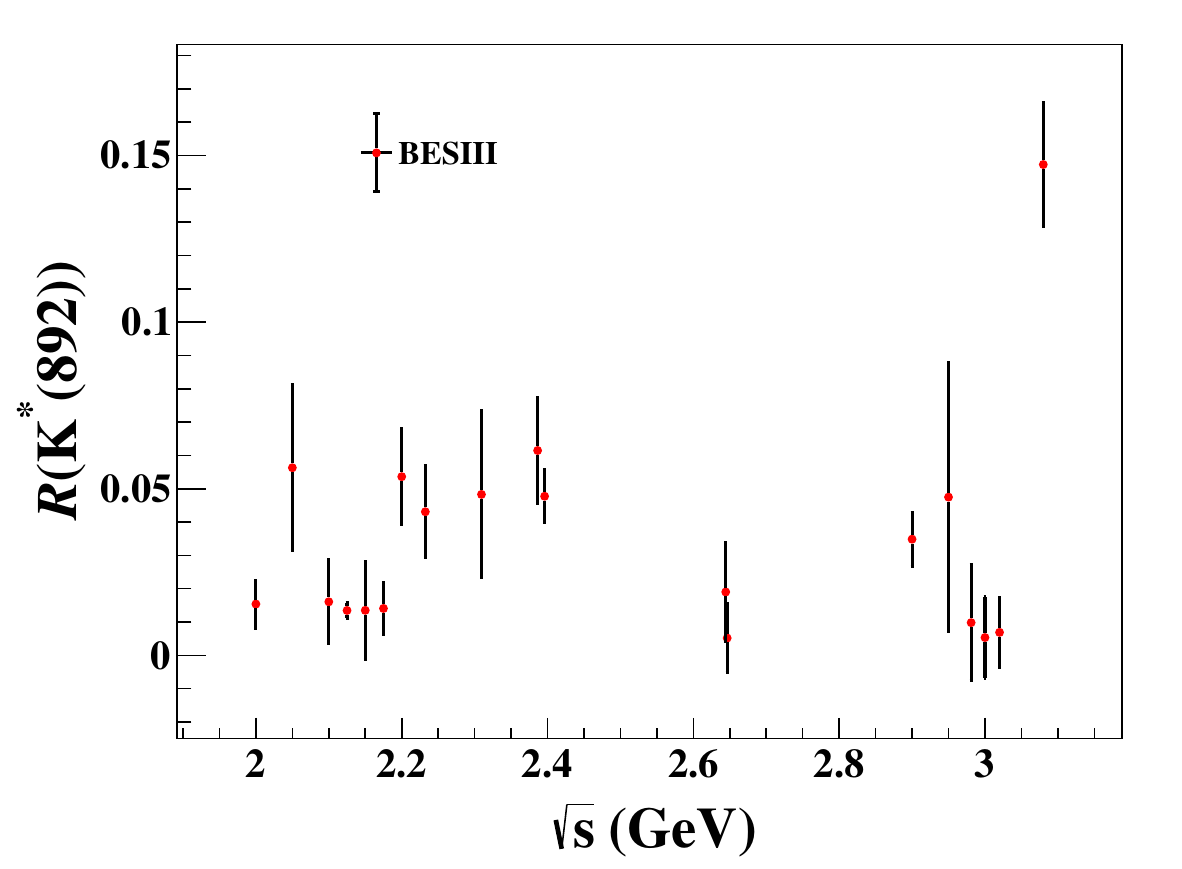}
        \put(20,60){(a)}
        \end{overpic}
        \begin{overpic}[width=0.48\textwidth]{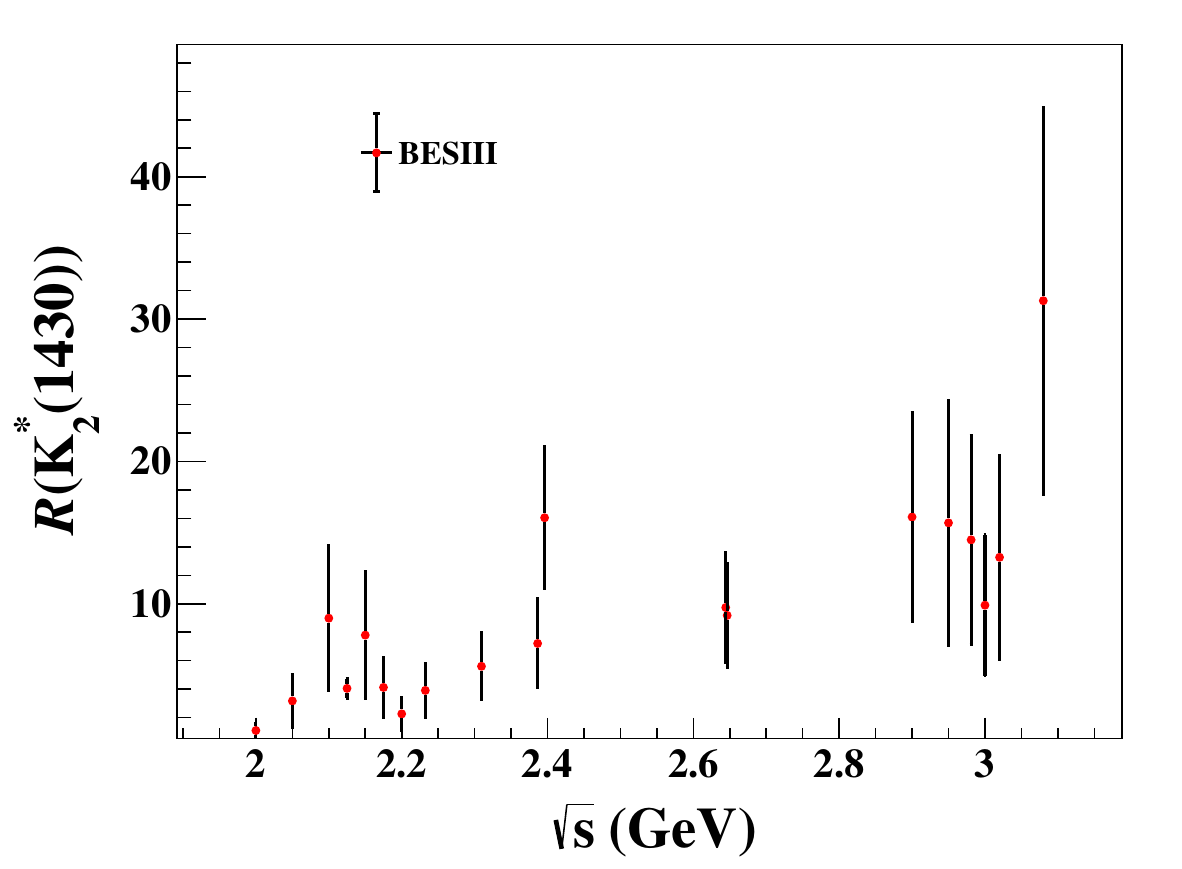}
        \put(20,60){(b)}
        \end{overpic}
        \caption{Relative ratio distributions for (a) $K^{*}(892)$ and (b) $K^{*}_{2}(1430)$.}
        \label{ratio:k892}
    \end{figure}

\noindent The corresponding results for the relative ratio are summarized in
figure~\ref{ratio:k892}.

    \begin{table}[t]
        \centering
            \begin{tabular}{cr@{\,}c@{\,}lr@{\,}c@{\,}lcccr@{\,}c@{\,}c@{\,}c@{\,}c@{\,}c@{\,}l}\\ \hline
            $\sqrt{s}$ (GeV)  &\multicolumn{3}{c}{$N_{\rm{sig}}$}  &\multicolumn{3}{c}{$\mathcal{L}$ (pb$^{-1}$)} &$\epsilon$ &1+$\delta$ &$\frac{1}{|1-\Pi|^{2}}$ &\multicolumn{7}{c}{$\sigma$ (pb)}\\ \hline
            2.0000              &$880.6$ & $\pm$ & $32.3$       &$10.1$ & $\pm$ & $0.1$         &15.3\%     &1.22   &1.037      &$662.7$ & $\pm$ & $24.3$ & $\pm$ & $17.7$ & $\pm$ &$18.0$ \\
            2.0500              &$308.4$ & $\pm$ & $19.2$       &$3.34$ & $\pm$ & $0.03$       &15.8\%  &1.20   &1.038          &$682.1$ & $\pm$ & $42.5$ & $\pm$ & $18.3$ & $\pm$ &$26.9$ \\
            2.1000              &$941.7$ & $\pm$ & $33.8$       &$12.2$ & $\pm$ & $0.1$     &16.4\%     &1.18   &1.039          &$557.8$ & $\pm$ & $20.0$ & $\pm$ & $14.9$ & $\pm$ &$15.0$ \\
            2.1250              &$8175.0$ & $\pm$ & $98.9$      &$108$ & $\pm$ & $ 1$           &16.6\%     &1.17   &1.039      &$545.4$ & $\pm$ & $6.6$ & $\pm$ & $14.6$ & $\pm$ &$7.2$ \\
            2.1500              &$228.1$ & $\pm$ & $16.8$       &$2.84$ & $\pm$ & $0.02$       &17.1\%  &1.13   &1.040          &$582.2$ & $\pm$ & $42.9$ & $\pm$ & $15.6$ & $\pm$ &$23.5$ \\
            2.1750              &$678.7$ & $\pm$ & $28.4$       &$10.6$ & $\pm$ & $0.1$        &16.5\%  &1.23   &1.040          &$444.5$ & $\pm$ & $18.6$ & $\pm$ & $11.9$ & $\pm$ &$16.0$ \\
            2.2000              &$772.8$ & $\pm$ & $30.5$       &$13.7$ & $\pm$ & $0.1$        &16.2\%  &1.25   &1.040          &$391.2$ & $\pm$ & $15.4$ & $\pm$ & $10.5$ & $\pm$ &$12.2$ \\
            2.2324              &$594.1$ & $\pm$ & $26.7$       &$11.9$ & $\pm$ & $0.1$        &16.0\%  &1.28   &1.041          &$342.9$ & $\pm$ & $15.4$ & $\pm$ & $9.2$ & $\pm$ &$11.1$ \\
            2.3094              &$846.1$ & $\pm$ & $32.8$       &$21.1$ & $\pm$ & $0.1$         &15.5\%     &1.31   &1.041      &$277.3$ & $\pm$ & $10.8$ & $\pm$ & $7.4$ & $\pm$ &$10.6$ \\
            2.3864              &$741.3$ & $\pm$ & $30.1$       &$22.5$ & $\pm$ & $0.2$        &15.5\%  &1.34   &1.041          &$224.2$ & $\pm$ & $9.1$ & $\pm$ & $6.0$ & $\pm$ &$7.9$ \\
            2.3960              &$2146.1$ & $\pm$ & $51.2$      &$66.9$ & $\pm$ & $0.5$        &15.2\%  &1.34   &1.041          &$220.4$ & $\pm$ & $5.3$ & $\pm$ & $5.9$ & $\pm$ &$4.6$ \\
            2.6444              &$595.2$ & $\pm$ & $27.7$      &$33.7$ & $\pm$ & $0.2$        &15.0\%     &1.45   &1.039        &$114.4$ & $\pm$ & $5.3$ & $\pm$ & $3.1$ & $\pm$ &$3.2$ \\
            2.6464              &$615.4$ & $\pm$ & $27.8$      &$34.0$ & $\pm$ & $0.3$        &15.0\%     &1.45   &1.039        &$117.2$ & $\pm$ & $5.3$ & $\pm$ & $3.1$ & $\pm$ &$3.2$ \\
            2.9000              &$1100.7$ & $\pm$ & $37.1$      &$105$ & $\pm$ & $1$        &14.0\%     &1.65   &1.033          &$64.8$ & $\pm$ & $2.2$ & $\pm$ & $1.7$ & $\pm$ &$1.4$ \\
            2.9500              &$124.0$ & $\pm$ & $13.0$       &$15.9$ & $\pm$ & $0.1$        &13.3\%  &1.68   &1.029          &$49.6$ & $\pm$ & $5.2$ & $\pm$ & $1.3$ & $\pm$ &$2.3$ \\
            2.9810              &$146.0$ & $\pm$ & $13.8$      &$16.1$ & $\pm$ & $0.1$          &13.9\%     &1.66   &1.025      &$56.1$ & $\pm$ & $5.3$ & $\pm$ & $1.5$ & $\pm$ &$2.1$ \\
            3.0000              &$144.5$ & $\pm$ & $13.9$       &$15.9$ & $\pm$ & $0.1$        &13.7\%  &1.67   &1.021          &$56.6$ & $\pm$ & $5.4$ & $\pm$ & $1.5$ & $\pm$ &$2.7$ \\
            3.0200              &$143.8$ & $\pm$ & $13.6$       &$17.3$ & $\pm$ & $0.1$        &13.7\%  &1.71   &1.014          &$51.1$ & $\pm$ & $4.8$ & $\pm$ & $1.4$ & $\pm$ &$1.9$ \\
            3.0800              &$963.8$ & $\pm$ & $35.3$       &$126$ & $\pm$ & $1$            &12.8\%     &1.83   &0.915      &$52.3$ & $\pm$ & $1.9$ & $\pm$ & $1.4$ & $\pm$ &$1.2$ \\
            \hline
            \end{tabular}
            \caption{The measured Born cross sections for
            $e^{+}e^{-}\to\Ks\Kl\pi^{0}$, where the first
            uncertainties are statistical, the second ones are
            systematics from table~\ref{tab:corrected} and the third ones are model uncertainties.}
        \label{tab:cs3}
        \end{table}

        \begin{table}[t]
            \centering
                \begin{tabular}{cr@{\,}c@{\,}lr@{\,}c@{\,}lcccr@{\,}c@{\,}c@{\,}c@{\,}c@{\,}c@{\,}l}\\ \hline
            $\sqrt{s}$ (GeV)  &\multicolumn{3}{c}{$N_{\rm{sig}}$}  &\multicolumn{3}{c}{$\mathcal{L}$ (pb$^{-1}$)} &$\epsilon$ &1+$\delta$ &$\frac{1}{|1-\Pi|^{2}}$ &\multicolumn{7}{c}{$\sigma$ (pb)}\\ \hline
                2.0000 				&$845.4$&$\pm$&$45.3$   	&$10.1$&$\pm$&$0.1$      	&17.0\% 	&1.07   &1.037	 	&$1942.5$&$\pm$&$104.0$&$\pm$& $52.0$ & $\pm$ &$63.3$\\
                2.0500 				&$225.4$&$\pm$&$22.8$   	&$3.34$&$\pm$&$0.03$      &17.2\% 	&1.08   &1.038	 	    &$1533.4$&$\pm$&$155.4$&$\pm$& $41.0$ & $\pm$ &$71.1$\\
                2.1000 				&$772.9$&$\pm$&$52.7$   	&$12.2$&$\pm$&$0.1$   	&17.6\% 	&1.07	&1.039 		    &$1420.4$&$\pm$&$96.9$&$\pm$& $38.0$ & $\pm$ &$46.1$\\
                2.1250 				&$6530.7$&$\pm$&$120.8$   &$108$&$\pm$&$1$       	&17.7\% 	&1.05   &1.039	 	    &$1376.2$&$\pm$&$25.5$&$\pm$& $36.8$ & $\pm$ &$19.3$\\  
                2.1500 				&$200.6$&$\pm$&$24.4$   	&$2.84$&$\pm$&$0.02$      &17.1\% 	&1.10   &1.040	 	    &$1585.8$&$\pm$&$193.1$&$\pm$& $42.5$ & $\pm$ &$63.0$\\
                2.1750 				&$516.1$&$\pm$&$31.7$   	&$10.6$&$\pm$&$0.1$       &17.3\% 	&1.13   &1.040		    &$1055.1$&$\pm$&$64.8$&$\pm$& $28.2$ & $\pm$ &$33.1$\\
                2.2000 				&$575.5$&$\pm$&$33.6$   	&$13.7$&$\pm$&$0.1$       &16.9\% 	&1.14  	&1.040 	 	    &$918.5$&$\pm$&$53.6$&$\pm$& $24.6$ & $\pm$ &$31.0$\\ 
                2.2324 				&$472.0$&$\pm$&$26.9$   	&$11.9$&$\pm$&$0.1$       &16.9\% 	&1.16   &1.041	 	    &$851.5$&$\pm$&$48.5$&$\pm$& $22.8$ & $\pm$ &$27.0$\\ 
                2.3094 				&$637.6$&$\pm$&$32.5$   	&$21.1$&$\pm$&$0.1$      	&16.4\% 	&1.20   &1.041   	&$648.1$&$\pm$&$33.0$&$\pm$& $17.3$ & $\pm$ &$22.4$\\ 
                2.3864 				&$595.6$&$\pm$&$31.5$   	&$22.5$&$\pm$&$0.2$       &16.2\% 	&1.23   &1.041 	 	    &$559.5$&$\pm$&$29.6$&$\pm$& $15.0$ & $\pm$ &$17.3$\\
                2.3960 				&$1845.8$&$\pm$&$53.0$   	&$66.9$&$\pm$&$0.5$       &16.4\% 	&1.23   &1.041   	    &$578.4$&$\pm$&$16.6$&$\pm$& $15.5$ & $\pm$ &$11.4$\\
                2.6444 				&$438.0$&$\pm$&$25.5$   	&$33.7$&$\pm$&$0.2$       &15.5\% 	&1.35   &1.039   	    &$262.0$&$\pm$&$15.3$&$\pm$& $7.0$ & $\pm$ &$8.1$\\ 
                2.6464 				&$452.9$&$\pm$&$25.9$   	&$34.0$&$\pm$&$0.3$       &15.5\% 	&1.35   &1.039   	    &$268.5$&$\pm$&$15.3$&$\pm$& $7.2$ & $\pm$ &$8.2$\\ 
                2.9000 				&$799.7$&$\pm$&$35.6$   	&$105$&$\pm$&$1$      	&14.9\% 	&1.50   &1.033   	    &$145.8$&$\pm$&$6.5$&$\pm$& $3.9$ & $\pm$ &$3.9$\\ 
                2.9500 				&$94.8$&$\pm$&$11.0$  	&$15.9$&$\pm$&$0.1$       &14.3\% 	&1.54   &1.029   	        &$115.5$&$\pm$&$13.4$&$\pm$& $3.1$ & $\pm$ &$6.9$\\ 
                2.9810 				&$111.6$&$\pm$&$11.7$     &$16.1$&$\pm$&$0.1$ 		&14.4\%   	&1.56   &1.025		    &$132.0$&$\pm$&$13.8$&$\pm$& $3.5$ & $\pm$ &$7.3$\\  
                3.0000 				&$108.0$&$\pm$&$11.5$   	&$15.9$&$\pm$&$0.1$       &14.2\% 	&1.58   &1.021   	    &$130.5$&$\pm$&$13.9$&$\pm$& $3.5$ & $\pm$ &$6.9$\\ 
                3.0200 				&$104.3$&$\pm$&$10.8$   	&$17.3$&$\pm$&$0.1$       &14.2\% 	&1.59   &1.014   	    &$115.7$&$\pm$&$12.0$&$\pm$& $3.1$ & $\pm$ &$6.2$\\ 
                3.0800 				&$695.9$&$\pm$&$33.4$   	&$126$&$\pm$&$1$       	&13.3\% 	&1.73   &0.915   	    &$115.1$&$\pm$&$5.5$&$\pm$& $3.1$ & $\pm$ &$2.4$\\ 
                \hline
                \end{tabular}
                \caption{The measured Born cross sections for
                $e^{+}e^{-}\to\Kstar\Kbar$, where the first
                uncertainties are statistical, the second ones are
            systematics from table~\ref{tab:corrected} and the third ones are model uncertainties.}
            \label{tab:csk892}
            \end{table}

            \begin{table}[t]
                \centering
                    \begin{tabular}{cr@{\,}c@{\,}lr@{\,}c@{\,}lcccr@{\,}c@{\,}c@{\,}c@{\,}c@{\,}c@{\,}l}\\ \hline
            $\sqrt{s}$ (GeV)  &\multicolumn{3}{c}{$N_{\rm{sig}}$}  &\multicolumn{3}{c}{$\mathcal{L}$ (pb$^{-1}$)} &$\epsilon$ &1+$\delta$ &$\frac{1}{|1-\Pi|^{2}}$ &\multicolumn{7}{c}{$\sigma$ (pb)}\\ \hline
                    2.0000              &$80.1$&$\pm$&$28.2$     &$10.1$&$\pm$&$0.1$       &20.0\%     &1.06   &1.037         &$315.2$&$\pm$&$111.2$&$\pm$& $8.4$ & $\pm$ &$11.2$\\
                    2.0500              &$11.1$&$\pm$&$5.0$      &$3.34$&$\pm$&$0.03$      &20.4\%     &1.07   &1.038         &$127.5$&$\pm$&$58.2$&$\pm$& $3.4$ & $\pm$ &$6.0$\\
                    2.1000              &$27.3$&$\pm$&$13.7$     &$12.2$&$\pm$&$0.1$       &20.7\%     &1.07   &1.039         &$84.6$&$\pm$&$42.4$&$\pm$& $2.3$ & $\pm$ &$3.6$\\
                    2.1250              &$606.2$&$\pm$&$68.3$    &$108$&$\pm$&$1$         &20.5\%     &1.08   &1.039         &$214.2$&$\pm$&$24.1$&$\pm$& $5.7$ & $\pm$ &$5.2$\\  
                    2.1500              &$9.7$&$\pm$&$5.1$       &$2.84$&$\pm$&$0.02$      &23.4\%     &0.95   &1.040         &$129.6$&$\pm$&$68.1$&$\pm$& $3.5$ & $\pm$ &$6.1$\\
                    2.1750              &$59.1$&$\pm$&$28.6$     &$10.6$&$\pm$&$0.1$       &26.1\%     &0.88   &1.040         &$204.1$&$\pm$&$98.7$&$\pm$& $5.5$ & $\pm$ &$7.5$\\
                    2.2000              &$113.1$&$\pm$&$54.7$    &$13.7$&$\pm$&$0.1$       &25.8\%     &0.96   &1.040         &$304.0$&$\pm$&$147.0$&$\pm$& $8.1$ & $\pm$ &$12.2$\\ 
                    2.2324              &$55.8$&$\pm$&$25.2$     &$11.9$&$\pm$&$0.1$       &23.8\%     &1.03   &1.041         &$171.9$&$\pm$&$77.8$&$\pm$& $4.6$ & $\pm$ &$7.5$\\ 
                    2.3094              &$54.5$&$\pm$&$21.6$     &$21.1$&$\pm$&$0.1$       &22.3\%     &1.04   &1.041         &$94.3$&$\pm$&$37.4$&$\pm$& $2.5$ & $\pm$ &$4.3$\\ 
                    2.3864              &$28.0$&$\pm$&$11.5$     &$22.5$&$\pm$&$0.2$       &22.3\%     &1.07   &1.041         &$45.3$&$\pm$&$18.7$&$\pm$& $1.2$ & $\pm$ &$1.8$\\
                    2.3960              &$41.5$&$\pm$&$12.3$     &$66.9$&$\pm$&$0.5$       &21.7\%     &1.07   &1.041         &$22.3$&$\pm$&$6.6$&$\pm$& $0.6$ & $\pm$ &$0.7$\\
                    2.6444              &$21.5$&$\pm$&$8.2$      &$33.7$&$\pm$&$0.2$       &22.0\%     &1.07   &1.039         &$22.9$&$\pm$&$8.8$&$\pm$& $0.6$ & $\pm$ &$0.9$\\ 
                    2.6464              &$22.3$&$\pm$&$8.4$      &$34.0$&$\pm$&$0.3$       &22.0\%     &1.07   &1.039         &$23.5$&$\pm$&$8.9$&$\pm$& $0.6$ & $\pm$ &$0.9$\\ 
                    2.9000              &$20.4$&$\pm$&$9.1$      &$105$&$\pm$&$1$          &22.7\%     &1.08   &1.033         &$6.7$&$\pm$&$3.0$&$\pm$& $0.2$ & $\pm$ &$0.2$\\ 
                    2.9500              &$2.2$&$\pm$&$1.1$       &$15.9$&$\pm$&$0.1$       &22.5\%     &1.08   &1.029         &$4.8$&$\pm$&$2.4$&$\pm$& $0.1$ & $\pm$ &$0.3$\\ 
                    2.9810              &$2.5$&$\pm$&$1.1$       &$16.1$&$\pm$&$0.1$       &22.9\%     &1.08   &1.025         &$5.3$&$\pm$&$2.4$&$\pm$& $0.1$ & $\pm$ &$0.3$\\  
                    3.0000              &$4.7$&$\pm$&$2.2$       &$15.9$&$\pm$&$0.1$       &22.8\%     &1.08   &1.021         &$10.3$&$\pm$&$4.7$&$\pm$& $0.3$ & $\pm$ &$0.6$\\ 
                    3.0200              &$2.2$&$\pm$&$1.1$       &$17.3$&$\pm$&$0.1$       &23.0\%     &1.08   &1.014         &$4.5$&$\pm$&$2.1$&$\pm$& $0.1$ & $\pm$ &$0.2$\\ 
                    3.0800              &$8.7$&$\pm$&$3.7$       &$126$&$\pm$&$1$          &23.6\%     &1.08   &0.915         &$2.6$&$\pm$&$1.1$&$\pm$& $0.1$ & $\pm$ &$0.1$\\ 
                    \hline
                    \end{tabular}
                    \caption{The measured Born cross sections for
                    $e^{+}e^{-}\to\KstarT\Kbar$, where the first
                    uncertainties are statistical, the second ones are
            systematics from table~\ref{tab:corrected} and the third ones are model uncertainties.}
                \label{tab:csk2}
                \end{table}

\section{Systematic uncertainties} Two categories of systematic
uncertainties are considered in the measurement of the Born cross
sections. The first category includes systematic sources not associated with the PWA fit that are evaluated as follows:
    
\begin{enumerate} 

\item The uncertainty associated with the integrated luminosity is
$1\%$ and estimated by using large angle Bhabha
events~\cite{besiii_lumin}.

\item The uncertainty concerning $\Ks$ reconstruction is studied with
control samples of $J/\psi \to K_{S}^{0}K^{\pm}\pi^{\mp}$ and
$J/\psi\to \phi K_{S}^{0}K^{\pm}\pi^{\mp}$.  The result shows that the
difference in efficiency between data and MC simulation is 1\% per
$K_{S}^{0}$~\cite{bes_ks_sys}.

\item The uncertainty of the requirement on the number of charged
tracks~($N_{\rm{charge}}$) is estimated with a control sample of \mbox{$J/\psi \to
K^{0}_{S}K^{0}_{L}\pi^{0}$}. The difference in efficiency between data
and MC simulation with and without this requirement is taken as the
uncertainty.

\item The uncertainty concerning photon detection efficiency is
studied with a control sample of $e^{+}e^{-}\to
K^{+}K^{-}\pi^{+}\pi^{-}\pi^{0}$~\cite{besiii_pi0}. The result shows
that the difference in detection efficiency between data and MC
simulation is 1\% per photon.

\item The uncertainty related to the kinematic fit is studied with a
control sample of \mbox{$J/\psi \to K^{0}_{S}K^{0}_{L}\pi^{0}$}. The
difference in efficiency between data and MC simulation with and without
the kinematic fit is taken as the uncertainty.


\item The uncertainty of the VP and ISR correction factors~(Rad) is obtained
with the accuracy of the radiation function, which is about
0.5\%~\cite{VP}, and has an additional contribution from the cross
section line shape, which is estimated by varying the model parameters
of the fit to the cross section. All parameters are randomly varied
within their uncertainties, and the resulting parametrization of the
line shape is used to recalculate $(1+\delta)\epsilon$ and the
corresponding cross section. This procedure is repeated one thousand
times, and the standard deviation of the resulting cross sections is
taken as the systematic uncertainty. The systematic uncertainty
associated with the VP and ISR correction factor
is evaluated as the quadratic sum of contributions from 
the QED theory and line shape parametrization~\cite{besiii_kk}.

\item The uncertainty associated with the BFs from the
PDG~\cite{PDG} is $0.08\%$, including both
$\mathcal{B}(\Ks\to\pi^{+}\pi^{-})=(69.20\pm0.05)\%$ and
$\mathcal{B}(\pi^{0}\to\gamma\gamma)=(98.823\pm0.034)\%$.

\item The uncertainty caused by the $M(\pi^{+}\pi^{-})$ fit~(Fit) includes
the descriptions of signal shape and background shape. The
nominal MC-simulated shape convolved with a Gaussian function is
replaced by a MC-simulated shape convolved with a Crystal Ball function, and the nominal background
shape is replaced by a second-order polynomial function, and the
differences with the nominal results are taken as the
uncertainties. The uncertainties from above sources are added in
quadrature and taken as the total uncertainty from the
$M(\pi^{+}\pi^{-})$ fit.

\end{enumerate}

    The second category of uncertainties includes those associated with the PWA fit that are evaluated as follows:

\begin{enumerate}

\item The uncertainty from the fit parameters~(FPar) is estimated by the
standard deviation of re-calculated efficiencies derived from one
thousand groups of randomly generated fit parameters using a
correlated multi-variable Gaussian function.


\item The uncertainty related to the resonance parameters~(Par) is estimated
by performing alternative fits shifting the world-average parameter value 
by its error from the PDG~\cite{PDG}.

\item The uncertainty concerning the extra additional resonances~(Extra) is estimated by performing alternative fits with all components whose significances are greater than 3$\sigma$. In the alternative fit  $K^{*}(1410)^{0}\to K^{0}\pi^{0}$ is added for the data sample at $\sqrt{s}=2.125$ GeV, $K^{*}(1410)^{0}\to K^{0}\pi^{0}$ and $\rho(1450)\to K^{0}\bar{K}^{0}$ are added at $\sqrt{s}=2.396$ GeV, $\rho(1450)/\rho(1700) \to K^{0}\bar{K}^{0}$ and $K^{*}(1410)^{0}/K^{*}_{3}(1780)^{0}\to K^{0}\pi^{0}$ are added at $\sqrt{s}=2.900$ GeV.

\item The uncertainty of the background estimation in the PWA fit~(Bkg) is
estimated by using only the lower or higher sideband.

\item The uncertainty from the Blatt-Weisskopf barrier factor~(BWf) is estimated by varying the radius of the centrifugal barrier from 0.7 to 1.0 fm.
        
\end{enumerate}

Assuming all the sources of systematic uncertainties as independent,
the total systematic uncertainty is obtained by adding them in
quadrature. The 100\% correlated uncertainties for the Born cross
sections of $e^{+}e^{-}\to\Ks\Kl\pi^{0}$, $\Kstar\Kbar$ and
$\KstarT\Kbar$ are listed in table~\ref{tab:corrected}. The other uncorrelated and total
systematic uncertainties are listed in
tables~\ref{tab:sys3}-\ref{tab:sysk2}.

    \begin{table}[t]
        \centering
        \begin{tabular}{lc}
            \hline
            Source &Uncertainty(\%)\\ \hline
            Luminosity &1.00 \\
            $\Ks$ reconstruction &1.00 \\
            Requirement on $N_{\rm{charge}}$ &0.70 \\
            Photon reconstruction &2.00 \\
            Kinematic fit &0.70 \\
            BF &0.08 \\ \hline
            Total &2.63 \\ \hline
        \end{tabular}
        \caption{The 100\% correlated systematic uncertainties for the Born cross section of $e^{+}e^{-}\to\Ks\Kl\pi^{0}$.}
        \label{tab:corrected}
        \end{table}

    \begin{table}[t]
        \centering
        \begin{tabular}{ccccccccc}
            \hline
            $\sqrt{s}$~(GeV) &$\rm{Rad}$ &Fit &FPar &Par &Extra &Bkg &BWf &Total \\ \hline
            2.0000 &0.50  &0.91  &2.40   &0.70 &0.41 &0.26 &0.21  &2.76\\ 
            2.0500 &0.50  &2.13  &2.50   &2.11 &0.40 &0.30 &0.21  &3.97\\ 
            2.1000 &0.50  &0.36  &2.40   &1.02 &0.40 &0.30 &0.17  &2.73\\ 
            2.1250 &0.50  &0.94  &0.80   &0.22 &0.38 &0.07 &0.13  &1.41\\ 
            2.1500 &0.50  &2.83  &2.60   &1.10 &0.43 &0.30 &0.31  &4.07\\ 
            2.1750 &0.50  &1.14  &3.00   &1.53 &0.40 &0.24 &0.40  &3.64\\ 
            2.2000 &0.50  &1.39  &2.50   &1.11 &0.41 &0.30 &0.22  &3.16\\        
            2.2324 &0.50  &2.11  &2.20   &1.02 &0.32 &0.30 &0.17  &3.29\\ 
            2.3094 &0.50  &2.71  &2.40   &1.10 &0.31 &0.28 &0.22  &3.84\\
            2.3864 &0.50  &1.44  &3.00   &1.10 &0.30 &0.30 &0.21  &3.57\\
            2.3960 &0.50  &1.36  &1.40   &0.60 &0.30 &0.15 &0.21  &2.14\\
            2.6444 &0.50  &2.12  &1.41   &1.02 &0.30 &0.18 &0.40  &2.84\\
            2.6464 &0.50  &2.05  &1.40   &1.02 &0.30 &0.18 &0.40  &2.78\\
            2.9000 &0.50  &1.50  &1.30   &0.68 &0.29 &0.20 &0.13  &2.19\\
            2.9500 &0.51  &3.20  &2.57   &2.10 &0.33 &0.30 &0.31  &4.67\\
            2.9810 &0.50  &2.50  &2.20   &1.69 &0.31 &0.30 &0.30  &3.80\\
            3.0000 &0.50  &3.75  &2.20   &2.06 &0.30 &0.28 &0.25  &4.86\\
            3.0200 &0.50  &2.58  &1.70   &1.91 &0.30 &0.30 &0.30  &3.70\\
            3.0800 &0.50  &0.98  &1.70   &1.02 &0.31 &0.20 &0.20  &2.30\\ \hline
        \end{tabular}
        \caption{Systematic uncertainties~(\%) for the Born cross section of $e^{+}e^{-}\to\Ks\Kl\pi^{0}$ at each c.m. energy associated with the ISR and VP correction factors~(Rad), the M($\pi^{+}\pi^{-})$ fit~(Fit), the fit parameters in PWA~(FPar), the resonance parameters~(Par), the extra additional resonances~(Extra), the background estimation~(Bkg) and the Blatt-Weisskopf barrier factor~(BWf).}
        \label{tab:sys3}
    \end{table}

    \begin{table}[t]
        \centering
        \begin{tabular}{cccccccc}
            \hline
            $\sqrt{s}$~(GeV) &$\rm{Rad}$ &Fit   &Par   &Extra   &Bkg     &BWf     &Total \\ \hline
            2.0000 &0.50  &0.91   &1.90 &0.71 &1.43 &1.91  &3.30\\ 
            2.0500 &0.50  &2.13   &1.22 &0.70 &2.32 &3.10  &4.67\\ 
            2.1000 &0.50  &0.36   &1.71 &0.70 &1.72 &2.01  &3.29\\ 
            2.1250 &0.50  &0.94   &0.21 &0.68 &0.20 &0.73  &1.49\\ 
            2.1500 &0.50  &2.83   &1.23 &0.73 &1.30 &2.01  &4.00\\ 
            2.1750 &0.50  &1.14   &1.01 &0.70 &1.12 &2.40  &3.17\\ 
            2.2000 &0.50  &1.39   &1.32 &0.71 &1.52 &2.22  &3.42\\        
            2.2324 &0.50  &2.11   &1.10 &0.52 &1.15 &1.67  &3.21\\ 
            2.3094 &0.50  &2.71   &1.05 &0.51 &1.09 &1.42  &3.49\\
            2.3864 &0.50  &1.44   &1.04 &0.51 &1.14 &2.20  &3.13\\
            2.3960 &0.50  &1.36   &0.53 &0.50 &0.50 &1.13  &2.03\\
            2.6444 &0.50  &2.12   &0.72 &0.51 &1.20 &1.70  &3.14\\
            2.6464 &0.50  &2.05   &0.72 &0.51 &1.17 &1.70  &3.08\\
            2.9000 &0.50  &1.50   &0.38 &0.69 &1.57 &1.32  &2.71\\
            2.9500 &0.51  &3.20   &3.01 &0.73 &2.31 &3.18  &5.96\\
            2.9810 &0.50  &2.50   &3.01 &0.71 &2.11 &3.17  &5.53\\
            3.0000 &0.50  &3.75   &1.53 &0.70 &2.01 &2.68  &5.33\\
            3.0200 &0.50  &2.58   &1.22 &0.70 &1.42 &4.20  &5.34\\
            3.0800 &0.50  &0.98   &0.50 &0.71 &0.97 &1.27  &2.13\\ \hline
        \end{tabular}
        \caption{Systematic uncertainties~(\%) for the Born cross section of $e^{+}e^{-}\to\Kstar\Kbar$ at each c.m. energy associated with the ISR and VP correction factors~(Rad), the M($\pi^{+}\pi^{-})$ fit~(Fit), the fit parameters in PWA~(FPar), the resonance parameters~(Par), the extra additional resonances~(Extra), the background estimation~(Bkg) and the Blatt-Weisskopf barrier factor~(BWf).}
        \label{tab:sysk892}
    \end{table}

    \begin{table}[t]
        \centering
        \begin{tabular}{cccccccc}
            \hline
            $\sqrt{s}$~(GeV) &$\rm{Rad}$ &Fit &Par &Extra &Bkg &BWf &Total \\ \hline
            2.0000 &0.50  &0.91   &2.02 &1.00 &1.50 &2.07  &3.56\\ 
            2.0500 &0.50  &2.13   &2.10 &1.00 &2.70 &2.21  &4.73\\ 
            2.1000 &0.50  &0.36   &2.27 &1.00 &2.58 &2.21  &4.25\\ 
            2.1250 &0.50  &0.94   &1.01 &1.00 &1.05 &1.29  &2.43\\ 
            2.1500 &0.50  &2.83   &1.82 &1.00 &2.22 &2.08  &4.67\\ 
            2.1750 &0.50  &1.14   &2.10 &1.00 &1.50 &2.10  &3.69\\ 
            2.2000 &0.50  &1.39   &2.10 &1.00 &1.92 &2.21  &4.02\\        
            2.2324 &0.50  &2.11   &2.12 &1.31 &2.00 &2.01  &4.35\\ 
            2.3094 &0.50  &2.71   &2.10 &1.30 &1.50 &2.10  &4.51\\
            2.3864 &0.50  &1.44   &2.10 &1.30 &1.62 &2.10  &3.93\\
            2.3960 &0.50  &1.36   &1.49 &1.31 &1.31 &1.70  &3.26\\
            2.6444 &0.50  &2.12   &1.70 &1.31 &1.67 &2.00  &4.02\\
            2.6464 &0.50  &2.05   &1.70 &1.31 &1.70 &2.00  &3.99\\
            2.9000 &0.50  &1.50   &1.38 &1.50 &1.20 &1.92  &3.43\\
            2.9500 &0.51  &3.20   &2.50 &1.53 &2.00 &2.20  &5.28\\
            2.9810 &0.50  &2.50   &2.72 &1.52 &2.03 &2.11  &4.98\\
            3.0000 &0.50  &3.75   &2.51 &1.52 &2.11 &2.20  &5.68\\
            3.0200 &0.50  &2.58   &2.51 &1.52 &2.12 &2.00  &4.90\\
            3.0800 &0.50  &0.98   &1.70 &1.51 &1.20 &1.92  &3.39\\ \hline
        \end{tabular}
        \caption{Systematic uncertainties~(\%) for the Born cross section of $e^{+}e^{-}\to\KstarT\Kbar$ at each c.m. energy associated with the ISR and VP correction factors~(Rad), the M($\pi^{+}\pi^{-})$ fit~(Fit), the resonance parameters~(Par), the extra additional resonances~(Extra), the background estimation~(Bkg) and the Blatt-Weisskopf barrier factor~(BWf).}
        \label{tab:sysk2}
    \end{table}

\section{Fit to the lineshape} 

\begin{figure}[t]
        \centering
        \begin{overpic}[width=0.48\textwidth]{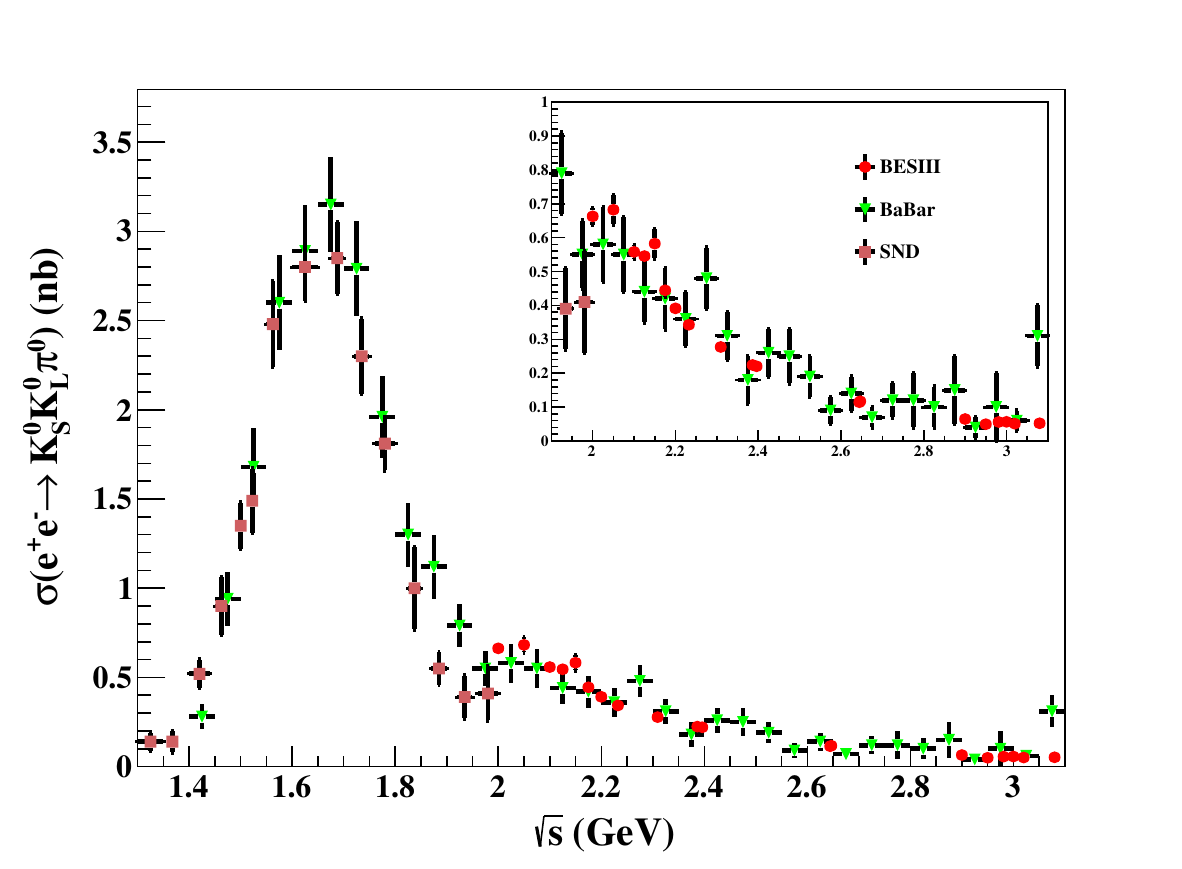}
        \put(15,60){(a)}
        \end{overpic}
        \begin{overpic}[width=0.48\textwidth]{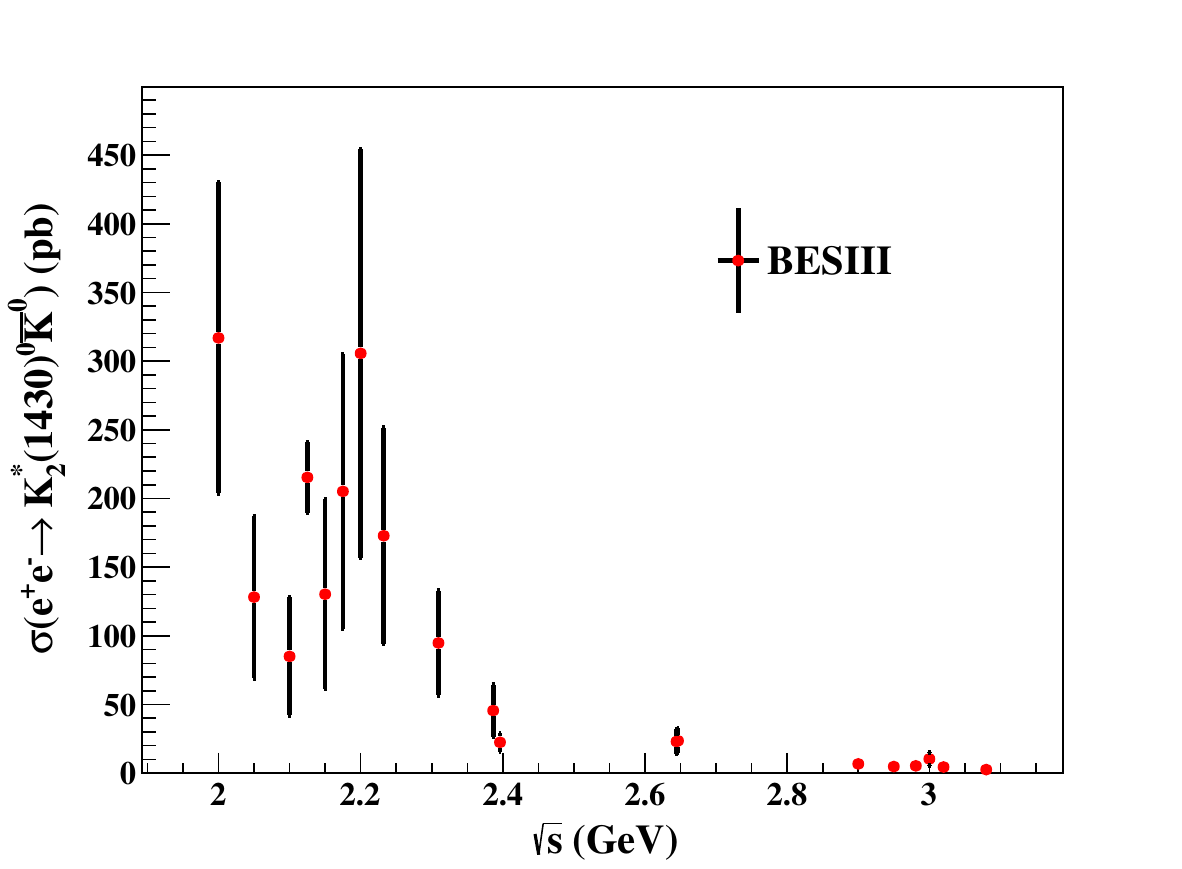}
        \put(15,60){(b)}
        \end{overpic}
        \caption{The Born cross sections for (a) the process $e^{+}e^{-}\to \Ks\Kl\pi^{0}$ and (b) the process $e^{+}e^{-}\to \KstarT\Kbar$. The red dots are the measured results from BESIII, where errors include both statistical and systematic uncertainties. The green triangles and brown squares are the results from BaBar and SND, respectively.}
        \label{fig:3bodylineshape}
    \end{figure}

\begin{figure}[t]
        \centering
        \begin{overpic}[width=0.48\textwidth]{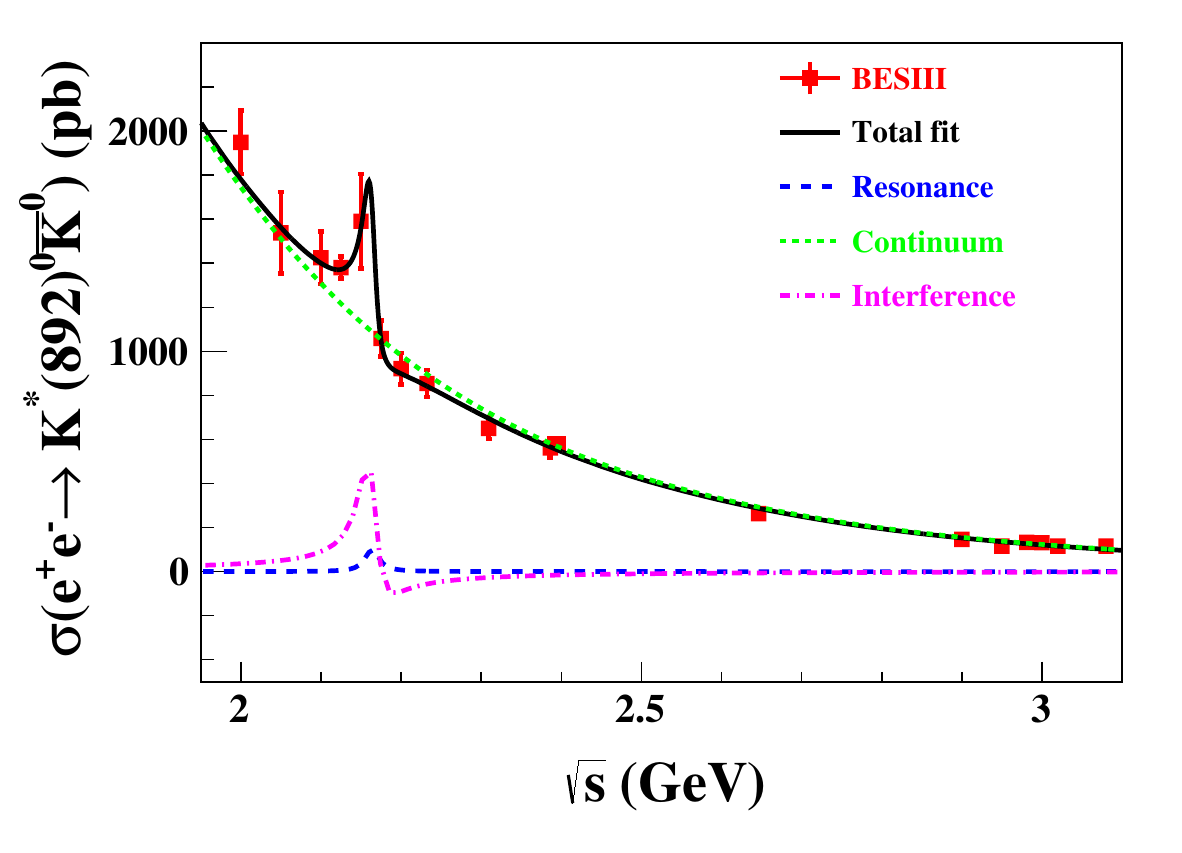}
        \put(35,60){(a)}
        \end{overpic}
        \begin{overpic}[width=0.48\textwidth]{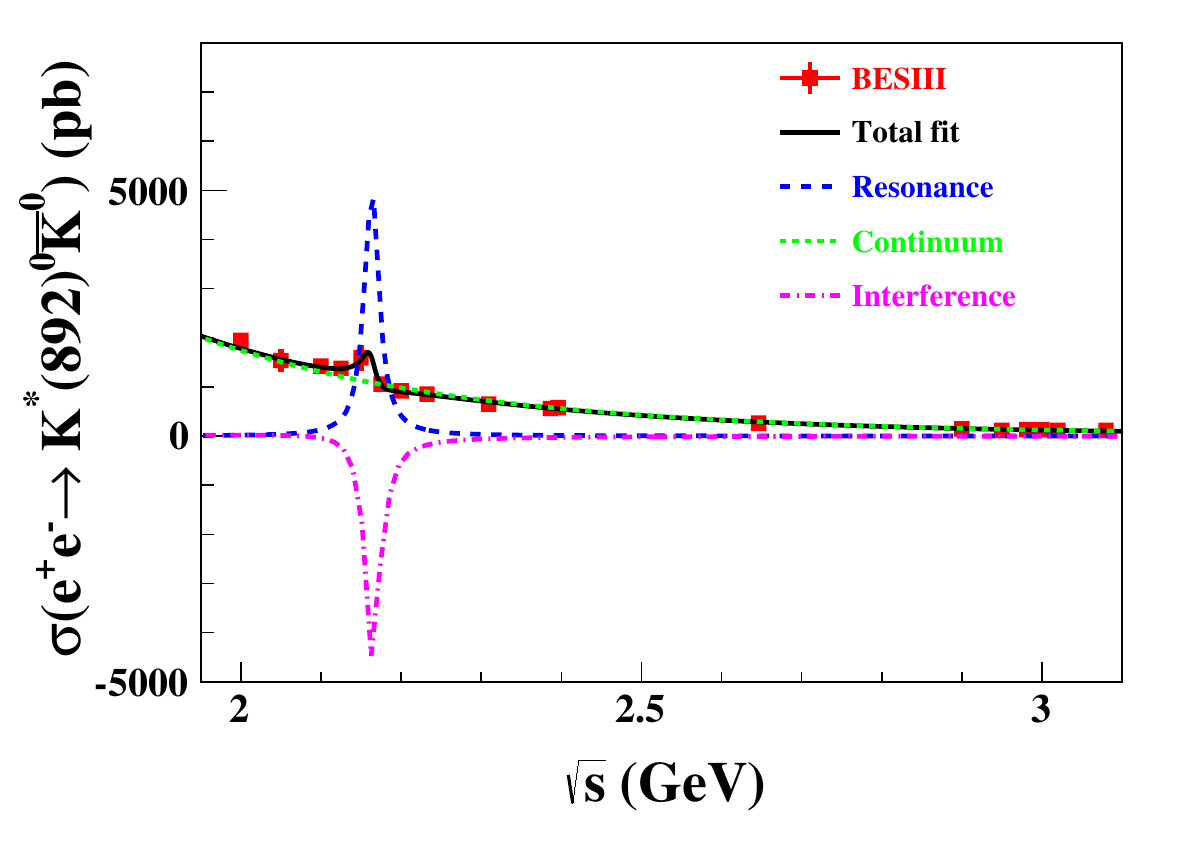}
        \put(35,60){(b)}
        \end{overpic}
        \flushleft
        \caption{The Born cross section and fit curves for
        $e^{+}e^{-}\to\Kstar\Kbar$, (a) and (b), corresponding to the
        two solutions in table~\ref{tab:fitresult}. Rectangles with
        error bars are BESIII data, where errors include both
        statistical and systematic uncertainties. The solid black
        curves represent the total fit result, the dashed blue
        curves for the resonance and the dashed green curves for the
        continuum component, and the dash-dotted pink
        curves for the interference between the resonance and
        continuum components.}
        \label{fig:k892lineshape}
    \end{figure}

The Born cross sections for the process $e^{+}e^{-}\to\Ks\Kl\pi^{0}$
are shown in figure~\ref{fig:3bodylineshape}(a). The results are
consistent with the previous results from BaBar. The Born
cross sections for the intermediate process
$e^{+}e^{-}\to\KstarT\Kbar$ and $e^{+}e^{-}\to\Kstar\Kbar$ are shown
in figures~\ref{fig:3bodylineshape}(b) and \ref{fig:k892lineshape},
respectively.

A $\chi^{2}$ fit, incorporating the correlated and uncorrelated
uncertainties among different energy points, is performed to determine
the resonance parameters for the Born cross sections of
$e^{+}e^{-}\to\Kstar\Kbar$. The fit probability density function is a
coherent sum of a continuum component and a resonant component. The
cross section is modeled as: 

\begin{equation}
\sigma=|\frac{\sqrt{12\pi\Gamma_Y\Gamma^{e^{+}e^{-}}_Y\mathcal{B}}}{M_{Y}^{2}-s-iM_{Y}\Gamma(\sqrt{s})}
\sqrt{\frac{P(\sqrt{s})}{P(M_Y)}}e^{i\phi_Y}+c_1
\frac{\sqrt{P(\sqrt{s})}}{s^{c_2}}|^{2}, \end{equation} 

\noindent where $M_Y$ and $\Gamma_Y$ are the mass and width of the resonance;
$\phi_Y$ is the relative phase between the continuum component and the
resonance; $\Gamma_{Y}^{e^{+}e^{-}}$ is its partial width to
$e^{+}e^{-}$;  $\mathcal{B}$ is the BF of $Y\to\Kstar\Kbar$; and 
$c_{1}$ and $c_{2}$ are additional parameters of the 
fit. $\Gamma(\sqrt{s})$ is defined as 
$\Gamma_Y(\frac{P(\sqrt{s})}{P(M_Y)})$, where
$P(\sqrt{s})=\int\sum_{m}|A_{\Kstar\Kbar}(m)|^2 d\Phi_3$ 
is the phase-space factor for the relative orbital angular momentum
$L=1$ of the process $e^{+}e^{-}\to\Kstar\Kbar\to\Ks\Kl\pi^{0}$ and
$\Phi_3$ is three-body phase space~\cite{PDG}. The amplitude $A$ is
the partial wave amplitude in the covariant Rarita-Schwinger tensor
formalism~\cite{zoubs} and is described as:
    \begin{equation}
        A_{\Kstar\Kbar}(m)=-Y^{\mu}(m)\epsilon_{\mu\nu\lambda\sigma}p^{\sigma}_{(K^{0}_{S}K^{0}_{L}\pi^{0})}\tilde{T}^{(1)\nu}_{K^{*}(892)^{0}\Kbar}\cdot f^{K^{*}(892)^{0}}_{K^{0}\pi^{0}}\cdot \tilde{T}^{(1)\lambda}_{K^{0}\pi^{0}},
    \end{equation}
    where $\tilde{T}$ is the covariant tensor, $f$ is the Breit-Wigner propagator, $\epsilon_{\mu\nu\lambda\sigma}$ is the Levi-Civita symbol, and the other operators can be found in Ref.~\cite{zoubs}.

In total, there are six free parameters in the fit:
$M_Y,\Gamma_Y,\phi_Y,c_1,c_2$ and the product of
$\Gamma_{Y}^{e^{+}e^{-}}\mathcal{B}$. The fit results are shown in
figure~\ref{fig:k892lineshape}, and the resonance parameters are listed
in table~\ref{tab:fitresult}. The fit has $\chi^{2}/\rm{ndf}=18.95/13$($p=0.125$)
with two solutions for the resonance with identical mass and width,
but different relative phase $\phi_Y$ and
$\Gamma_{Y}^{e^{+}e^{-}}\mathcal{B}$. The mass and width of the resonance
are $M_{Y}=(2164.1\pm9.6)~\rm{MeV}/c^{2}$ and
$\Gamma_{Y}=(32.4\pm21.1)~\rm{MeV}$, respectively, including all sources of uncertainties from cross section measurement. The significance of the resonance
is determined to be $3.2\sigma$ by comparing the change of
$\chi^{2}$~($\Delta\chi^{2}=18.27$) and the change of ndf~($\Delta
\rm{ndf}=4$)~between the nominal fit and the fit without the
resonance. As both solutions are mathematically equivalent, we do not prefer one
over the other, but list them in order of increasing interference fraction. Figure~\ref{fig:k892lineshape}(b) shows a large
interference between resonance and continuum components. The
uncertainties~(statistical and systematic) of the measured Born cross
sections have been included when fitting the line shape, determining
the resonance parameters and estimating the significance of the
resonance.

Besides the uncertainties of individual cross-section measurements,
the fit to the lineshape is also affected by the uncertainty
of the BEPCII c.m. energy and the description of the continuum. The
uncertainty of the c.m. energy calibration is estimated as $0.1\%$ and
is ignored in the determination of resonance
parameters~\cite{besiii_lumin}. To evaluate the systematic uncertainty
associated with the lineshape model, the continuum term $c_1
\frac{\sqrt{P(\sqrt{s})}}{s^{c_2}}$ is replaced with an exponential
function of the form $c_{0}\cdot e^{-p_{0}(\sqrt{s}-M_{th})}$, where
$c_0$ and $p_{0}$ are free parameters and
$M_{th}=m_{\Kstar}+m_{\Kbar}$ is the mass threshold for $\Kstar\Kbar$
production~\cite{PDG,besiii_kkpi}. The difference of the
parameters from the nominal results are taken as the systematic
uncertainties.

    \begin{table}[t]
        \centering
        \begin{tabular}{l|cc}  
        \hline
             Parameter      &Solution 1 &Solution 2\\\hline
    
             $M_{Y}$(MeV/$c^2$) &\multicolumn{2}{c}{$2164.7\pm9.1\pm3.1$} \\ 
             $\Gamma_{Y}$(MeV)  &\multicolumn{2}{c}{$32.4\pm21.0\pm1.8$}  \\ 
             $\Gamma^{e^+e^-}_{Y}\mathcal{B}$(eV) &$1.0\pm0.2\pm0.1$ &$73.6\pm4.4\pm2.0$  \\
             $\phi_Y$(rad)        &$2.5\pm0.5\pm0.1$ &$-1.7\pm0.1\pm0.1$  \\
             Significance       &\multicolumn{2}{c}{$3.2\sigma$}  \\
            \hline
        \end{tabular}
        \caption{Result of the fit to the $e^{+}e^{-}\to \Kstar \Kbar$ Born cross sections, where the first uncertainties originate from the cross section measurement and the second from the line shape fit methodology, respectively.}
        \label{tab:fitresult}
        \end{table}

\section{Summary} In summary, a partial wave analysis of the process
$e^{+}e^{-}\to\Ks\Kl\pi^{0}$ is performed for nineteen data samples
collected in the BESIII experiment with center-of-mass energies
ranging from 2.000 to 3.080 GeV corresponding to a total integrated
luminosity of 647 $\rm{pb}^{-1}$. The Born cross sections of the
process $e^{+}e^{-}\to\Ks\Kl\pi^{0}$, as well as those for the
intermediate processes $e^{+}e^{-}\to\Kstar\Kbar$ and $\KstarT\Kbar$
are measured by performing partial wave analysis on each data
sample individually, where the charge conjugated processes are also
included. The measured Born cross sections of the process
$e^{+}e^{-}\to\Ks\Kl\pi^{0}$ are consistent with earlier
results by BaBar~\cite{babar_ksklpi}, while
the precision is significantly improved. The Born cross section
lineshape of the process $e^{+}e^{-}\to\Kstar\Kbar$ hints at 
a resonant structure around 2.2 GeV with a significance of
$3.2\sigma$. A Breit-Wigner fit yields its mass
$M_Y=(2164.7\pm9.1\pm3.1)~{\rm{MeV}}/c^{2}$ and width
$\Gamma_{Y}=(32.4\pm21.0\pm1.8)~\rm{MeV}$. The resonance parameters,
especially the very narrow width, are very close to the BESIII results
measured through the $\phi\eta$ channel~\cite{bes_phieta} of the
$\phi(2170)$ meson~\cite{PDG}.
 
The ratio of Born cross section measurements
of the process $e^{+}e^{-}\to\Kstarc K^{-}$ to the process
\mbox{$e^{+}e^{-}\to\Kstar\Kbar$} is less than 0.2 and that for
$K_{2}^{*}(1430)$ is in the region of 0-40, where the statistical and
systematic uncertainties are included. If we apply the isospin
decomposition for the decay from isospin vector~($\rho^{*}$) or
isospin scalar~($\omega^{*},\phi^{*}$) state to the final state
$K^{*}\bar{K}$, the ratio of the yields in the neutral and charged $K^{*}\bar{K}$
should be 1. On the other hand, the electromagnetic interaction 
also contributes to the production of $e^{+}e^{-}\to
K^{*}\bar{K}$, and it does not require isospin conservation. 
Future experimental and theoretical studies are needed
to understand the observed phenomenon.

    \acknowledgments
    The BESIII Collaboration thanks the staff of BEPCII, the IHEP computing center and the supercomputing center of USTC for their strong support. This work is supported in part by National Key R\&D Program of China under Contracts Nos. 2020YFA0406400, 2020YFA0406300; National Natural Science Foundation of China (NSFC) under Contracts Nos. 11635010, 11735014, 11835012, 11935015, 11935016, 11935018, 11961141012, 12022510, 12025502, 12035009, 12035013, 12192260, 12192261, 12192262, 12192263, 12192264, 12192265, 12275320, 11625523, 11705192, 11950410506, 12061131003, 12105276, 12122509; the Chinese Academy of Sciences (CAS) Large-Scale Scientific Facility Program; the CAS Center for Excellence in Particle Physics (CCEPP); Joint Large-Scale Scientific Facility Funds of the NSFC and CAS under Contracts Nos. U1832207, U1732263, U1832103, U2032111; CAS Key Research Program of Frontier Sciences under Contracts Nos. QYZDJ-SSW-SLH003, QYZDJ-SSW-SLH040; 100 Talents Program of CAS; The Institute of Nuclear and Particle Physics (INPAC) and Shanghai Key Laboratory for Particle Physics and Cosmology; ERC under Contract No. 758462; European Union's Horizon 2020 research and innovation programme under Marie Sklodowska-Curie grant agreement under Contract No. 894790; German Research Foundation DFG under Contracts Nos. 443159800, 455635585, Collaborative Research Center CRC 1044, FOR5327, GRK 2149; Istituto Nazionale di Fisica Nucleare, Italy; Ministry of Development of Turkey under Contract No. DPT2006K-120470; National Research Foundation of Korea under Contract No. NRF-2022R1A2C1092335; National Science and Technology fund; National Science Research and Innovation Fund (NSRF) via the Program Management Unit for Human Resources \& Institutional Development, Research and Innovation under Contract No. B16F640076; Polish National Science Centre under Contract No. 2019/35/O/ST2/02907; The Royal Society, UK under Contracts Nos. DH140054, DH160214; The Swedish Research Council; U. S. Department of Energy under Contract No. DE-FG02-05ER41374.

    \providecommand{\href}[2]{#2}\begingroup\raggedright\endgroup
\newpage
\section*{The BESIII collaboration}
\addcontentsline{toc}{section}{The BESIII collaboration}
M.~Ablikim$^{1}$, M.~N.~Achasov$^{5,b}$, P.~Adlarson$^{75}$, X.~C.~Ai$^{81}$, R.~Aliberti$^{36}$, A.~Amoroso$^{74A,74C}$, M.~R.~An$^{40}$, Q.~An$^{71,58}$, Y.~Bai$^{57}$, O.~Bakina$^{37}$
% ...
\\
\vspace{0.2cm}
(BESIII Collaboration)\\
\vspace{0.2cm} {\it
$^{1}$ Institute of High Energy Physics, Beijing 100049, People's Republic of China\\
% ...
$^{5}$ Budker Institute of Nuclear Physics SB RAS (BINP), Novosibirsk 630090, Russia\\
% ...
$^{36}$ Johannes Gutenberg University of Mainz, Johann-Joachim-Becher-Weg 45, D-55099 Mainz, Germany\\
$^{37}$ Joint Institute for Nuclear Research, 141980 Dubna, Moscow region, Russia\\
% ...
$^{40}$ Liaoning Normal University, Dalian 116029, People's Republic of China\\
% ...
$^{57}$ Southeast University, Nanjing 211100, People's Republic of China\\
$^{58}$ State Key Laboratory of Particle Detection and Electronics, Beijing 100049, Hefei 230026, People's Republic of China\\
% ...
$^{71}$ University of Science and Technology of China, Hefei 230026, People's Republic of China\\
% ...
$^{74}$ University of Turin and INFN, (A)University of Turin, I-10125, Turin, Italy; (B)University of Eastern Piedmont, I-15121, Alessandria, Italy; (C)INFN, I-10125, Turin, Italy\\
$^{75}$ Uppsala University, Box 516, SE-75120 Uppsala, Sweden\\
% ...
$^{81}$ Zhengzhou University, Zhengzhou 450001, People's Republic of China\\

\vspace{0.2cm}
\noindent
% ...
$^{b}$ Also at the Novosibirsk State University, Novosibirsk, 630090, Russia\\
% ...
}
\end{document}